\documentclass[12pt]{article}
\pdfoutput=1

\usepackage[pdftex]{graphicx}
\usepackage{amssymb}
\usepackage{amsmath}
\usepackage{cite}

\usepackage{wrapfig}

\newcommand{\eq}[1]{\begin{equation} #1 \end{equation}}

\newcommand{\eqa}[1]{\begin{eqnarray} #1 \end{eqnarray}}

\newcommand{\av}[1]{\langle #1 \rangle}
\newcommand{\sss}{\scriptscriptstyle}

\newcommand{\heff}{\mathcal{H}_{\rm eff}}

\newcommand{\op}{\mathcal{O}}
\newcommand{\nn}{\nonumber}

\newcommand{\azeL}{{A_0^L}}
\newcommand{\azeR}{{A_0^R}}

\newcommand{\apeL}{{A_\perp^L}}
\newcommand{\apeR}{{A_\perp^R}}

\newcommand{\apaL}{{A_\|^L}}
\newcommand{\apaR}{{A_\|^R}}

\newcommand{\re}{{\rm Re}}
\newcommand{\im}{{\rm Im}}

\newcommand{\C}[1]{{\cal C}_{#1}}

\begin{document}

\begin{flushright}
{\small
LPT-ORSAY/15-14\\
QFET-2015-06\\
SI-HEP-2015-05 
}
\end{flushright}
$\ $
\vspace{-2mm}
\begin{center}
\Large \bf
Time dependence in $B\to V\ell\ell$ decays
\end{center}

\vspace{1mm}
\begin{center}
{\sf S\'ebastien Descotes-Genon$\,^{a}$ and Javier Virto$\,^{b}$
}
\end{center}

\begin{center}
{\em \small
$\,^{a}$ Laboratoire de Physique Th\'eorique, CNRS/Univ. Paris-Sud (UMR 8627)\\ 91405 Orsay Cedex, France\\[2mm]
$\,^{b}$ Theoretische Physik 1, Naturwissenschaftlich-Technische Fakult\"at,\\ Universit\"at Siegen, 57068 Siegen, Germany
}
\end{center}

\vspace{1mm}
\begin{abstract}\noindent
\vspace{-5mm}

We discuss the theory and phenomenology of $B_{d,s}\to V(\to M_1M_2)\ell\ell$ decays in the presence of neutral-meson mixing.
We derive expressions for the time-dependent angular distributions for
decays into CP eigenstates, and identify the relevant observables that can be extracted from
time-integrated and time-dependent analyses with or without tagging, with a focus on the
difference between measurements at $B$-factories and hadronic machines. We construct two observables
of interest, which we call $Q_8^-$ and $Q_9$, and which are theoretically clean at large recoil. We
compute these two observables in the Standard Model, and show that they have good potential for New Physics searches
by considering their sensitivity to benchmark New Physics scenarios consistent with current $b\to s\ell\ell$ data.  
These results apply to decays such as $B_d\to K^*(\to K_S\pi^0)\ell\ell$,
$B_s\to \phi (\to K_SK_L)\ell\ell$ and $B_s\to \phi(\to K^+K^-)\ell\ell$.

\end{abstract}


\section{Introduction}
\label{sec:intro}

Rare $B$ decays mediated by flavour-changing neutral currents constitute a unique playground to test the
Standard Model (SM) and search for New Physics (NP). Among these,
processes mediated by the quark-level $b\to s\ell\ell$ transition have received a great deal of attention following
a large programme of measurements at $B$-factories, LHCb and CMS:
branching ratios, CP asymmetries and angular distributions of $B\to K^{(*)}\mu^+\mu^-$
\cite{1301.1700,1304.6325,1308.1707,1308.3409,1403.8045,1406.6482,0904.0770,1108.0695,1204.3933,1403.8044,1408.0978} and
$B_s\to \phi\mu^+\mu^-$ \cite{1305.2168} decays, the branching ratio ${\cal B}(B_s\to \mu^+\mu^-)$
\cite{1307.5024,1307.5025,1411.4964}, and inclusive $B\to X_s\ell\ell$ observables \cite{1312.5364,1402.7134}.
Global fits to all $b\to s \gamma$ and $b\to s\ell\ell$ data have recently uncovered a pattern of tensions between theory and
experiment, triggered by the the analysis of the $B\to K^*\mu^+\mu^-$ angular distribution
\cite{1307.5683,1308.1501,1310.2478,1310.3887}, and followed by the measurement of the ratio
$R_K\equiv {\cal B}(B\to K\mu\mu)/{\cal B}(B\to Kee)$, which is consistent with
New Physics in $B\to K^*\mu\mu$ data and hints at lepton-flavour
non-universality~\cite{1408.1627,1408.4097,1410.4545,1411.3161,1411.0565,1412.1791,1412.7164}.
In this context, the study of new independent $b\to s\ell\ell$ observables is of great interest, as a
means to gather evidence for (or against) these tensions, and to fingerprint the resulting New Physics.

When dealing with decays of \emph{neutral} $B$ mesons, experimental observables are affected by particle-antiparticle mixing
(oscillations), with the decaying meson being either a $B$ or a $\bar B$ depending on the time of decay.
In the case of flavour-non-specific decays --such as decays into CP eigenstates-- in which the final state can arise from the decay of both $B$ and $\bar B$ mesons,
the mixing and decay processes interfere quantum-mechanically, leading to interesting phenomenological consequences
(for a review see for instance Refs.~\cite{0904.1869,Harrison:1998yr}). 
In particular, new observables arise compared to the case without mixing. These observables depend on the
experimental set-up ($B$-factory or hadronic machine), the presence of flavour tagging of the decaying $B$-meson,
and the possibility to perform time-dependent measurements (in contrast to the limitation to time-integrated observables).
In the case of $b\to s\ell\ell$ transitions, these effects have been so far taken into account
in the untagged time-integrated measurements of $B_s\to \phi\mu^+\mu^-$~\cite{0805.2525}
and $B_s\to \mu^+\mu^-$~\cite{1204.1737} at the LHC,
(see also the discussion in Ref.~\cite{1111.4882} in the case of $B_s\to VV$ decays).
Time-dependent angular analyses of $B_{d,s} \to V\ell\ell$ with tagging are much more challenging experimentally,
but might be reached at a high-luminosity flavour factory such as Belle-II \cite{1011.0352}.

In this paper we develop the theoretical framework and study the phenomenological advantages of time-dependent
$B_{d,s}\to V\ell\ell$ decays, spelling out the new observables that can be accessed, as well as the opportunities for
New Physics searches, both at $B$-factories and hadronic machines. While the formalism is valid for any decay of
the type $B_{d,s} \to V(\to M_1M_2)\ell\ell$ with $M_1M_2$ a CP eigenstate, we identify the following modes of interest: 
\begin{itemize}
\item $B_d\to K^{*0}(\to K_S\pi^0) \ell^+\ell^-$ 
\item $B_s\to \phi(\to K_S K_L) \ell^+\ell^-$ 
\item $B_s\to\phi(\to K^+K^-) \ell^+\ell^-$
\end{itemize}
As a summary of the main points to be discussed below, we shall see that:
\begin{itemize}

\item In the presence of mixing, the time-dependent angular distributions exhibit a new type
of angular coefficients, $h_i$ and $s_i$, apart from the usual coefficients accessible from
flavour-specific decays, $J_i$ and $\bar J_i$, c.f. Eqs.~(\ref{eq:J+Jt}),(\ref{eq:J-Jt}).

\item Time-integrated CP-averaged rates and CP-asymmetries, as measured at hadronic
machines, are affected by mixing effects in two ways, c.f. Eqs.~(\ref{eq:<J+Jt>Had}),(\ref{eq:<J-Jt>Had}):
1) The terms with $J_i\pm \bar J_i$
are multiplied by the factors $1/(1-y^2)$ and $1/(1+x^2)$ respectively, with
$y=\Delta\Gamma/(2\Gamma)$ and $x=\Delta m/\Gamma$.
2) New contributions proportional to the coefficients $h_i$ and $s_i$ arise. At $B$-factories only
the first type of corrections appear, and time-integrated quantities are independent of the coefficients
$s_i$ and $h_i$, c.f. Eqs.(\ref{eq:<J+Jt>Bfac}),(\ref{eq:<J-Jt>Bfac}).

\item We identify $s_8$ and $s_9$ as new observables of interest, which can be extracted
most conveniently from a time-dependent analysis with flavour tagging. Theoretically clean observables
can be built from $s_8$ and $s_9$; two such observables are $Q_8^-$ and $Q_9$, which are
clean at large hadronic recoil. These observables contain independent information, not accessible from
flavour-specific decays. Such observables could be studied, for instance, at a high-luminosity flavour
factory, where a separation between $B$ and $\bar B$ samples would be possible together with
a study of time dependence of the decay process.

\item The observables $Q_8^-$ and $Q_9$ can be predicted in the Standard Model with small uncertainties
(see Fig.~\ref{figQ8Q9}). In particular, $Q_9$ measures right-handed currents: in the case of a $b\to s$ transition,
$Q_9^{\rm SM}\simeq -\cos(\phi_q - 2\beta_s)$
to a very good precision, with $\phi_q$ the mixing angle of the $B_q$ system. In addition, these observables are very sensitive to
New Physics scenarios consistent with current $b\to s\gamma$ and $b\to s\ell\ell$ data, such as
models with $Z'$ bosons with vector and/or axial couplings to fermions.

\end{itemize}

The structure of this article is the following. We begin in Section~\ref{sec:2} with a discussion on time-dependent angular distributions: In Section~\ref{sec:2.1} we review the basic facts
of $B\to V\ell\ell$ decays without mixing. In Section~\ref{sec:2.2} we address the CP parities
associated to transversity amplitudes for $B\to V\ell\ell$ decays into CP eigenstates. In
Section~\ref{sec:2.4} we derive the expressions for the time-dependent angular distributions,
and identify the new angular observables $h_i(s)$ and $s_i(s)$ that arise in the presence of mixing, 
demonstrating in Section~\ref{sec:sym} that $s_{5,6s,8,9}(s)$ contain independent information not accessible from the angular distribution of flavour-specific decays. In Section~\ref{sec:3} we discuss in detail the two types of observables that can be obtained from the distributions in the presence of mixing:
time-integrated (Section~\ref{sec:t-int}) and time-dependent (Section~\ref{sec:Q8Q9}) observables.
We also define the observables $Q_8^-$ and $Q_9$, which are form-factor-independent at large recoil, and
we provide simplified expressions at the leading order of the effective theory in this limit.
Standard Model predictions for these observables and New Physics opportunities
are discussed in Section~\ref{sec:4}. Finally, we conclude in Section~\ref{sec:conc}.
Some details are relegated to the appendices. In Appendix~\ref{app1} we discuss the kinematics
of CP-conjugated $B\to V(\to M_1M_2)\ell\ell$ decays in terms of momentum invariants and the
different conventions for the kinematic angles that appear in the angular distributions.
In Appendix~\ref{appCP} we recall the determination of the CP parity for the different transversity amplitudes.
In Appendix~\ref{app2} we collect the expressions for the
coefficients $h_i(s)$ and $s_i(s)$ in terms of transversity amplitudes.

\section{Time-dependent angular distributions}
\label{sec:2}

\subsection{$B\to V\ell\ell$ decays without mixing}
\label{sec:2.1}

We first recall a few elements of the analysis of the exclusive $b\to s\ell\ell$ decays  of the type
$B\to V(\to M_1 M_2) \ell\ell$.
In this subsection, we consider a situation where no mixing occurs, and where the $M_1M_2$ state is not (necessarily)
a CP-eigenstate.

This process is described by the usual effective Hamiltonian, with  SM operators plus (potentially) NP operators with chirality flip, scalar or tensor structure \cite{0811.1214,1212.2263}:
\eq{
\heff = \frac{4 G_F}{\sqrt{2}}
\bigg[
\lambda_u\,\C1 \op_1^u + \lambda_c\,\C1 \op_1^c
-\lambda_t\sum_{i\in I} \C{i} \op_{i}
\bigg]\ ,
}
where $\lambda_q = V_{qb} V_{qs}^*$ and $I = \{ 3,4,5,6, 8, 7, 7', 9,9',10,10',S,S',P,P',T,T' \}$.
The operators $\op_{1,..,6}$ and $\op_8$ are hadronic operators of the type
$(\bar s \Gamma b)(\bar q \Gamma' q)$ and
$(\bar s \sigma^{\mu\nu} T_a P_R b) G_{\mu\nu}^a$ respectively \cite{9612313},
and contribute to $b\to s \ell\ell$ processes through a loop coupled to an electromagnetic current
(via $b\to s\gamma^*\to s\ell\ell$). These operators are not likely to receive significant contributions
from NP, as these would show up in non-leptonic $B$ decay amplitudes\footnote{New Physics contributions at the
$\sim 10\%$ level to the operators $\op_1$, $\op_2$ is not excluded \cite{1412.1446}. However, this should have a
small impact on $b\to s\ell\ell$ where the effect of semileptonic operators dominates.}.
The operators
$\op_{7^{(\prime)},9^{(\prime)},10^{(\prime)},S^{(\prime)},P^{(\prime)},T^{(\prime)}}$ are given by:
\begin{align}
{\cal O}_{7^{(\prime)}} &= \frac{e}{(4\pi)^2} m_b [\bar{s} \sigma^{\mu\nu} P_{R(L)} b] F_{\mu\nu} \;, &
{\op}_{S^{(\prime)}} &= \frac{e^2}{(4\pi)^2} [\bar{s} P_{R(L)} b][\bar{\ell}\,\ell]\;, \nonumber
\\
\label{effops}
{\cal O}_{9^{(\prime)}} &=  \frac{e^2}{(4\pi)^2} [\bar{s} \gamma^\mu P_{L(R)} b] [\bar{\ell} \gamma_\mu \ell] \;, &
{\op}_{P^{(\prime)}} &= \frac{e^2}{(4\pi)^2} [\bar{s} P_{R(L)} b][\bar{\ell}\gamma_5\ell]\;, \\ \nonumber
{\cal O}_{10^{(\prime)}} &=  \frac{e^2}{(4\pi)^2} [\bar{s} \gamma^\mu P_{L(R)} b] [\bar{\ell} \gamma_\mu \gamma_5 \ell]\;, &
{\op}_{T^{(\prime)}} &=  \frac{e^2}{(4\pi)^2}
[\bar{s} \sigma_{\mu\nu}P_{R(L)} b][\bar{\ell}\sigma^{\mu\nu}P_{R(L)}\ell]\;,
\end{align}
with $\sigma^{\mu\nu}=i[\gamma^\mu,\gamma^\nu]/2$ and $P_{L,R}=(1\mp\gamma_5)/2$.
In the SM, and at a scale $\mu_b=\op(m_b)$, the only non-negligible Wilson coefficients regarding the
the operators in Eq.~(\ref{effops}) are $\C7^\text{SM}(\mu_b)\simeq -0.3$,
$\C9^\text{SM}(\mu_b)\simeq 4$ and $\C{10}^\text{SM}(\mu_b)\simeq -4$ (see Table~\ref{tab:Ci});
but all might be affected by NP. Contributions to $B\to V\ell\ell$ from electromagnetic dipole operators
$\op_{7^{(\prime)}}$ are (like hadronic contributions) of the type $b\to s\gamma^*\to s\ell\ell$. Contributions from
semileptonic operators $\op_{9^{(\prime)},10^{(\prime)},S^{(\prime)},P^{(\prime)},T^{(\prime)}}$ are factorizable and their matrix elements
can be written as
\eq{
\av{V\ell\ell| \op_\text{sl} |B} = \av{V|\Gamma^M |B} \av{\ell\ell|\Gamma'_M|0}\ ,
}
where $M$ denotes a collection of Lorentz indices. It is clear that all hadronic, dipole, and
semileptonic contributions can be recast as decays of the form
\eq{
B\to V(\to M_1M_2) N(\to \ell^+\ell^-)\ ,
\label{eq:BVN}
}
where $N$ has the quantum numbers of a boson, whose coupling pattern is determined by the
operators arising in the effective Hamiltonian.
In the SM, the structure of $\op_7,\op_9,\op_{10}$ shows that $N$ are spin-1 particles, coupling to both left- and right-handed fermions. This is in agreement with the presence of $\gamma^*$ and $Z$ penguin contributions, but it is also able to reproduce the contribution from box diagrams involving two $W$ bosons and a neutrino ($(V-A)(V-A)$ structure in the SM). In an extension of the SM yielding scalar (tensor) operators, one should add $N$ bosons with spin 0 (spin 2 respectively). 

We will work under the following assumptions, inspired by the situation in the SM and in its most usual extensions
\begin{itemize}
\item CP might be violated in the decay $B\to V N$, but it is conserved in the decay $N\to \ell^+\ell^-$.
\item $N$ can have spin 0 or spin 1, but not spin 2 (no tensor currents in the effective Hamiltonian).
\end{itemize}

It proves useful to analyse such decays in terms of transversity amplitudes.
Let us call 
\begin{equation}
M_{mn}=\epsilon^{*\mu}_V(m)\epsilon^{*\nu}_N(n) M_{\mu\nu} 
\end{equation}
the helicity amplitudes for this decay, where  $m$ and $n$ denote the polarisations of the meson $V$ and the virtual boson $N$ decaying into the dilepton pair, respectively. 

If $N$ has spin 1, as the initial decaying particle has spin 0, the only combination of helicity amplitudes allowed are
$(m,n)=(0,0),(+1,+1),(-1,-1),(0,t)$, where $t$ denotes the timelike polarisation.
One can then define the transversity amplitudes~\cite{9907386,0502060,0811.1214}
\begin{equation}
A_{\perp}=\frac{M_{+1,+1}-M_{-1,-1}}{\sqrt{2}}\qquad A_{||}=\frac{M_{+1,+1}+M_{-1,-1}}{\sqrt{2}}\qquad A_0=M_{0,0}\qquad A_t=M_{0,t}\ .
\end{equation}
The spin-1 $N$ particle couples to the lepton pair either through $\bar\ell\gamma_\mu P_L \ell$ or
$\bar\ell\gamma_\mu P_R \ell$, and we can further separate left- from right-handed components in the amplitudes:
$A_{0}^{L}$, $A_{0}^{R}$, $A_{||}^{L}$, $A_{||}^{R}$, $A_{\perp}^{L}$, $A_{\perp}^{R}$.
On the other hand, due to current conservation and the structure of the time-like polarisation $\epsilon^{*\nu}_N(t)\propto (p_{\ell^+}+p_{\ell^-})^\nu$, one can see that $A_t$ corresponds to a pure axial coupling to the lepton pair,  vanishing in the massless limit.
In the case where $N$ is spin 0, the only combination of helicity amplitudes allowed is $(m,n)=(0,0)$. The effect of a spin-0 particle with a pseudoscalar coupling to leptons can be absorbed into $A_t$, whereas a scalar coupling requires a new amplitude, called $A_S$.

The spin-summed differential decay distribution is given by \cite{9907386,0205287}
\eqa{\label{dist}
\frac{d^4\Gamma(B\to V(\to M_1M_2) \ell^+\ell^-)}{ds\,d\!\cos\theta_M\,d\!\cos\theta_l\,d\phi}&=&\frac9{32\pi} \bigg[
J_{1s} \sin^2\theta_M + J_{1c} \cos^2\theta_M + J_{2s} \sin^2\theta_M\cos 2\theta_l\nn\\[1.5mm]
&&\hspace{-4.2cm} + J_{2c} \cos^2\theta_M \cos 2\theta_l
+ J_3 \sin^2\theta_M \sin^2\theta_l \cos 2\phi + J_4 \sin 2\theta_M \sin 2\theta_l \cos\phi\nn\\[1.5mm]
&&\hspace{-4.2cm}+ J_5 \sin 2\theta_M \sin\theta_l \cos\phi 
+ J_{6s} \sin^2\theta_M\cos\theta_l +  {J_{6c} \cos^2\theta_M}  \cos\theta_l\\[1.5mm]    
&&\hspace{-4.2cm}+ J_7 \sin 2\theta_M \sin\theta_l \sin\phi  + J_8 \sin 2\theta_M \sin 2\theta_l \sin\phi 
+ J_9 \sin^2\theta_M \sin^2\theta_l \sin 2\phi \bigg]\,,\nn
}
in terms of the invariant mass of the lepton pair $s$, and three kinematical angles
$\theta_\ell,\theta_M,\phi$ (see Appendix~\ref{app1}). The coefficients of the distribution $J_i(s)$ contain interferences
of the form ${\rm Re}[A_XA_Y^*]$ and ${\rm Im}[A_XA_Y^*]$ between the eight
transversity amplitudes:
\begin{equation}
A_{0}^{L},\ A_{0}^{R},\ A_{||}^{L},\ A_{||}^{R},\ A_{\perp}^{L},\ A_{\perp}^{R},\ A_{t},\ A_S\ ,
\end{equation}
and are given by
\allowdisplaybreaks{
\eqa{
J_{1s}  & = & \frac{(2+\beta_\ell^2)}{4} \left[|\apeL|^2 + |\apaL|^2 +|\apeR|^2 + |\apaR|^2 \right]
+ \frac{4 m_\ell^2}{s} \re\left(\apeL\apeR^* + \apaL\apaR^*\right)\,,\nn\\[1mm]
J_{1c}  & = &  |\azeL|^2 +|\azeR|^2  + \frac{4m_\ell^2}{s} \left[|A_t|^2 + 2\re(\azeL^{}\azeR^*) \right] + \beta_\ell^2\, |A_S|^2 \,,\nn\\[1mm]
J_{2s} & = & \frac{ \beta_\ell^2}{4}\left[ |\apeL|^2+ |\apaL|^2 + |\apeR|^2+ |\apaR|^2\right],
\hspace{0.92cm}    J_{2c}  = - \beta_\ell^2\left[|\azeL|^2 + |\azeR|^2 \right]\,,\nn\\[1mm]
J_3 & = & \frac{1}{2}\beta_\ell^2\left[ |\apeL|^2 - |\apaL|^2  + |\apeR|^2 - |\apaR|^2\right],
\qquad   J_4  = \frac{1}{\sqrt{2}}\beta_\ell^2\left[\re (\azeL\apaL^* + \azeR\apaR^* )\right],\nn \\[1mm]
J_5 & = & \sqrt{2}\beta_\ell\,\Big[\re(\azeL\apeL^* - \azeR\apeR^* ) - \frac{m_\ell}{\sqrt{s}}\,
\re(\apaL A_S^*+ \apaR^* A_S) \Big]\,,\nn\\[1mm]
J_{6s} & = &  2\beta_\ell\left[\re (\apaL\apeL^* - \apaR\apeR^*) \right]\,,
\hspace{2.25cm} J_{6c} = 4\beta_\ell\, \frac{m_\ell}{\sqrt{s}}\, \re (\azeL A_S^*+ \azeR^* A_S)\,,\nn\\[1mm]
J_7 & = & \sqrt{2} \beta_\ell\, \Big[\im (\azeL\apaL^* - \azeR\apaR^* ) +
\frac{m_\ell}{\sqrt{s}}\, \im (\apeL A_S^* - \apeR^* A_S)) \Big]\,,\nn\\[1mm]
J_8 & = & \frac{1}{\sqrt{2}}\beta_\ell^2\left[\im(\azeL\apeL^* + \azeR\apeR^*)\right]\,,
\hspace{1.9cm} J_9 = \beta_\ell^2\left[\im (\apaL^{*}\apeL + \apaR^{*}\apeR)\right] \,,
\label{Js}}\\
}
where $\beta_\ell=\sqrt{1- 4m_\ell^2/s}$.
Similar expressions hold for the CP-conjugate decay $\bar{B}\to \bar{V}(\to \bar{M}_1 \bar{M}_2) \ell^+ \ell^-$,
with angular coefficients $\bar{J}_i$ involving amplitudes denoted by $\bar A_X$, and obtained from the $A_X$
by conjugating all weak phases~\footnote{This is opposite to the notation used in ref.~\cite{0811.1214} for $B$
and $\bar{B}$ decays, but in agreement with general discussions on CP-violation.}. The form of the angular distribution
for the CP-conjugated decay, however, depends on the way the kinematical variables
are defined. In the case in which the \emph{same} conventions are used irrespective of whether the
decaying meson is a $B$ or a $\bar B$, we have (see Appendix~\ref{app1}):
\begin{eqnarray}
\frac{d\Gamma[B\to V(\to M_1M_2) \ell^+\ell^-]}
{ds\  d\!\cos\theta_\ell\ d\!\cos\theta_M\ d\phi}&=&
\sum_i J_i(s) f_i(\theta_\ell,\theta_M,\phi)
\label{Gamma}\\
\frac{d\Gamma[\bar{B}\to \bar{V}(\to \bar{M}_1\bar{M}_2) \ell^+\ell^-]}
{ds\  d\!\cos\theta_\ell\ d\!\cos\theta_M\ d\phi}&=&
\sum_i \zeta_i\bar{J}_i(s) f_i(\theta_\ell,\theta_M,\phi)
\label{Gammabar}
\end{eqnarray}
where $f_i(\theta_\ell,\theta_M,\phi)$ are defined by Eq.~(\ref{dist}), and
\begin{equation}\label{eq:zetadef}
\zeta_i=1\quad{\rm for}\quad i=1s,1c,2s,2c,3,4,7\ ; \qquad
\zeta_i=-1\quad{\rm for}\quad i=5,6s,6c,8,9\ .
\end{equation}
We stress that this result arises just from the identification of kinematics of CP-conjugate decays, and
does not rely on any intrinsic CP-parity of the initial or final states involved.

\subsection{CP-parity of final states and decays into CP eigenstates}
\label{sec:2.2}

The separation into transversity amplitudes not only simplifies the analysis of the interference pattern, but also
provides amplitudes with final states possessing definite CP-parities\footnote{We emphasise that the term ``CP-parity"
makes reference to the final states and not to the amplitudes themselves, since the latter involve either a $B$ or a
$\bar{B}$,which are not CP-eigenstates.}.
In order to determine the CP-parities associated to the different transversity amplitudes we follow the analysis
of Ref.~\cite{Dunietz:1990cj}, where decays of the type $B\to M N$, with $M$,$N$ unstable particles, are considered.
The details of how to apply the results of Ref.~\cite{Dunietz:1990cj} to the $B\to V\ell\ell$ decays of interest
are provided in Appendix~\ref{appCP}; here we briefly summarize the main results.

We consider the decays $\bar B \to M_1M_2\ell^+\ell^-$ and $\bar B \to \bar M_1 \bar M_2\ell^+\ell^-$, such that
$M_1,M_2$ are either CP-eigenstates or CP-conjugates, and define the transversity amplitudes:
\eq{
\bar A_X \equiv A_X(\bar B \to \bar M_1 \bar M_2\ell^+\ell^-)\ ,\quad
\widetilde A_X \equiv A_X(\bar B \to M_1 M_2\ell^+\ell^-)\ ,
}
where $X=L0,R0,L\|,R\|,L\bot,R\bot,t,S$. These two sets of amplitudes are related by
\eq{
\widetilde A_X = \eta_X \bar A_X
}
where $\eta_X$ are the CP-parities associated to the different transversity amplitudes. We find that (see Appendix~\ref{appCP})
\eq{
\eta_X=\eta \quad \text{for}\quad  X=L0,L||,R0,R||,t \quad; \quad \eta_X=-\eta \quad \text{for}\quad X=L\perp,R\perp,S\ ,
\label{etaX}
}
where $\eta = 1$ if $M_1,M_2$ are CP conjugates (e.g. $K^+K^-$), and $\eta = -\eta(M_1) \eta(M_2)$ if $M_1,M_2$ are
CP eigenstates (e.g. $K_SK_L$). Here $\eta(M)$ denotes the intrinsic CP-parity of meson $M$. For the three processes
of interest mentioned in the introduction, the combination of intrinsic CP-parities leads always to $\eta=1$.

At this point we can classify the angular observables $J_i$ whether they combine amplitudes with identical or
opposite CP-parities, and whether they involve real or imaginary parts of interference terms:

\begin{itemize}

\item Real part with identical CP-parities: $i=1s,1c,2s,2c,3,4$.

\item Real part with opposite CP-parities: $i=5,6s,6c$.

\item Imaginary part with identical CP-parities: $i=7$. 

\item Imaginary part with opposite CP-parities: $i=8,9$. 

\end{itemize}

We note that the numbers $\zeta_i$ defined in Eq.~(\ref{eq:zetadef}) in a different context (identification of the kinematics between CP-conjugate decays) corresponds
to the product of the CP-parities of the amplitudes involved in the interference term $J_i$.\\

We now turn to the case of decays into CP eigenstates: $B\to f_{CP}$. In this context, it is useful to define two
different angular coefficients $\widetilde J_i$, $\bar J_i$ which are CP conjugates of $J_i$:

\begin{itemize}

\item the angular coefficients $\widetilde{J}_i$ formed by replacing $A_X$ by $\widetilde A_X\equiv A_X(\bar{B}\to f_{CP})$
(without CP-conjugation applied on $f_{CP}$), which appear naturally in the study of time evolution due to mixing,
where both $B$ and $\bar B$ decay into the same final state $f_{CP}$.

\item the angular coefficients $\bar{J}_i$, obtained by considering $\bar A_X\equiv A_X(\bar B \to \overline f_{CP})$ (with CP-conjugation applied to $f_{CP}$), which can be obtained from $A_X$ by changing the sign of all weak phases, and arise naturally when discussing CP violation from the theoretical point of view.

\end{itemize}
From the discussion above we have $\widetilde A_X = \eta_X \bar A_X$, with $\eta_X$ given in Eq.~(\ref{etaX}).
Plugging these amplitudes into the coefficients in Eq.~(\ref{Js}), we see that the two types of angular coefficients are
related through
\begin{equation}
\widetilde{J}_i=\zeta_i \bar{J}_i\ ,
\end{equation}
with $\zeta_i$ given in Eq.~(\ref{eq:zetadef}). In addition, in the limit of CP conservation, $J_i = \bar J_i$.

Since the final state is not self-tagging, an untagged measurement of the differential decay rate
(e.g. at LHCb, where the asymmetry production is tiny) yields essentially the CP-average
\begin{equation}
\frac{d\Gamma(B\to f_{CP})+d\Gamma(\bar{B}\to f_{CP})}
{ds\  d\!\cos\theta_\ell\ d\!\cos\theta_M\ d\phi}
=\sum_i [J_i+\widetilde{J}_i]  f_i(\theta_\ell,\theta_M,\phi)
=\sum_i [J_i+\zeta_i \bar{J}_i]  f_i(\theta_\ell,\theta_M,\phi)\ ,
\label{G+Gb}
\end{equation}
whereas the difference between the two decay rates (which can be measured only through flavour-tagging)
involves $J_i - \widetilde J_i = J_i-\zeta_i \bar{J}_i$,
\begin{equation}
\frac{d\Gamma(B\to f_{CP})-d\Gamma(\bar{B}\to f_{CP})}
{ds\  d\!\cos\theta_\ell\ d\!\cos\theta_M\ d\phi}
=\sum_i [J_i-\widetilde{J}_i]  f_i(\theta_\ell,\theta_M,\phi)
=\sum_i [J_i-\zeta_i \bar{J}_i]  f_i(\theta_\ell,\theta_M,\phi)\ .
\label{GmGb}
\end{equation}

We see that the convention chosen in Eqs.~(\ref{Gamma}),(\ref{Gammabar}) for flavour-tagging modes
allows one to treat on the same footing these modes and the modes with final CP-eigenstates, since the same combinations
of angular coefficients occur in both cases when one considers the CP-average or the CP-asymmetry in the decay rate.
Let us add that this results from a conventional identification between CP-conjugate decays in the case without mixing.
This freedom is not present in the presence of mixing where both decays result in the same final state,
which must always be described with the ``same" kinematic convention, in the sense of a convention that depends only on
the final state, without referring to the flavour of the decaying $B$ meson (see Appendix~\ref{app1}).

A slightly counter-intuitive consequence is that the CP-asymmetries for $J_i$ with
$i=5,6s,6c,8,9$ are measured in the CP-averaged rate, and vice-versa.
We also note that due to the interferences between different decay amplitudes,
only some of the $J_i-\bar{J}_i$ differences measure CP-violation in specific decay amplitudes
(i.e., $|\bar A|=|A|$, $i=1s,1c,2s,2c,3$) whereas the others measure relative phases between amplitudes
($i=4,5,6s,6c,7,8,9$), see Eq.~(\ref{Js}) .

\subsection{Angular distributions in the presence of mixing}
\label{sec:2.4}

In the case of $B$ decays into CP-eigenstates, where the final state can be produced both by the decay of $B$ or $\bar B$
mesons, the mixing and decay processes interfere, inducing a further time dependence in physical amplitudes
(see e.g. Ref.~\cite{0904.1869,Harrison:1998yr}). These time-dependent amplitudes are given by,
\begin{eqnarray}
A_X(t)&=&A_X(B(t)\to V(\to f_{CP})\to \ell^+\ell^-)=g_+(t) A_X + \frac{q}{p} g_-(t) \widetilde A_X\ ,\label{AXt}\\
\widetilde A_X(t)&=&A_X(\bar B(t)\to V(\to f_{CP}) \ell^+\ell^-)=\frac{p}{q}g_-(t) A_X + g_+(t) \widetilde A_X\ ,
\label{AbXt}
\end{eqnarray}
where the absence of the $t$ argument denotes the amplitudes at $t=0$, i.e. in the absence of mixing,
and we have introduced the usual time-evolution functions
\begin{eqnarray}
g_+(t)&=&e^{-imt}e^{-\Gamma t/2}\left[\cosh\frac{\Delta\Gamma t}{4}\cos\frac{\Delta m t}{2}-i\sinh\frac{\Delta\Gamma t}{4}\sin\frac{\Delta m t}{2}\right]\ ,\\
g_-(t)&=&e^{-imt}e^{-\Gamma t/2}\left[-\sinh\frac{\Delta\Gamma t}{4}\cos\frac{\Delta m t}{2}+i\cosh\frac{\Delta\Gamma t}{4}\sin\frac{\Delta m t}{2}\right]\ ,
\end{eqnarray}
with $\Delta m=M_H-M_L$ and $\Delta\Gamma=\Gamma_L-\Gamma_H$ (see Ref.~\cite{0904.1869}). The values of the
different mixing parameters for the three decays of interest are collected in Table~\ref{tab:params}.

In the presence of mixing, the coefficients of the angular distribution also become time-dependent, as they depend on the
time-dependent amplitudes in Eqs.~(\ref{AXt}),(\ref{AbXt}). This evolution can be simplified by noting that CP-violation
in $B_q-\bar B_q$ mixing is negligible for all practical purposes\footnote{The current world-averages
are $|q/p|_{B_d}=1.0007\pm 0.0009$ and $|q/p|_{B_d}=1.0038\pm 0.0021$ \cite{1412.7515}.},
and we will assume $|q/p|=1$, introducing the mixing angle $\phi$:
\begin{equation}
\frac{q}{p}=e^{i\phi}\ .
\end{equation}
This mixing angle is large in the case of the $B_d$ system but tiny for $B_s$, see Table~\ref{tab:params}.

\begin{table}
\begin{center}
\begin{tabular}{cccccccc}
Decay & $\eta$ & $\phi$ & $\sin\phi$ & $\cos\phi$ & $\Delta \Gamma$ & $x=\Delta m/\Gamma$ & $y=\Delta \Gamma/(2\Gamma)$\\
\hline
$B_d\to K^{*0}(\to K_S\pi^0) \ell^+\ell^-$  & 1 & $-2\beta$ & -0.7 & 0.7 & $\simeq 0$ & 0.77 & 0\\
$B_s\to \phi(\to K_L K_S) \ell^+\ell^-$ & 1 & $2\beta_s$ & 0.04 & 1 & $\neq 0$ & 27 & 0.06\\
$B_s\to \phi(\to K^+ K^-) \ell^+\ell^-$ & 1 & $2\beta_s$ & 0.04 & 1 & $\neq 0$ & 27 & 0.06\\
\hline
\end{tabular}
\caption{Parameters of the three decays of interest \cite{1412.7515}.}\label{tab:params}
\end{center}
\end{table}

The time-dependent angular coefficients are obtained by replacing time-independent amplitudes with time-dependent
ones in Eqs.~(\ref{Js}):
\eq{
J_i(t) = J_i \big(A_X\to A_X(t)\big)\ ,\quad
\widetilde J_i(t) = J_i \big(A_X\to \widetilde A_X(t)\big)\ .
\label{subst}}
We consider the combinations $J_i(t) \pm \widetilde J_i(t)$ appearing in the sum and difference of time-dependent
decay rates in Eqs.~(\ref{G+Gb}),~(\ref{GmGb}). From Eqs.~(\ref{AXt}), (\ref{AbXt}) and (\ref{subst}), we get
\begin{eqnarray}\label{eq:J+Jt}
J_i(t)+\widetilde J_i(t) &=&e^{-\Gamma t}\Big[(J_i + \widetilde J_i)\cosh(y\Gamma t) - h_i \sinh(y\Gamma t)\Big]\ ,\\[2mm]
J_i(t)-\widetilde J_i(t) &=&e^{-\Gamma t}\Big[(J_i - \widetilde J_i)\cos(x\Gamma t) - s_i \sin(x\Gamma t)\Big]\ ,
\label{eq:J-Jt}
\end{eqnarray}
where $x\equiv \Delta m/\Gamma$, $y\equiv \Delta \Gamma/(2\Gamma)$, and we have defined a new set of angular
coefficients $s_i,h_i$ related to the time-dependent angular distribution.
The coefficients $J_i$, $\widetilde J_i$ can already be determined from flavour-specific decays.
The explicit expressions
for $s_i$ and $h_i$ in terms of transversity amplitudes are collected in Appendix~\ref{app2}. 

Time-dependent angular distributions therefore contain potentially new information encoded in the new angular
observables $s_i$ and $h_i$. These pieces of information will be analysed in the rest of the paper. For the moment a
few comments are in order:

\begin{itemize}

\item The coefficients $h_i$ are very difficult to extract, since they are associated with $\sinh(y\Gamma t)$ with
$y$ very small. In particular, the time dependence of the untagged distribution (\ref{G+Gb}) provides essentially
no new information.

\item The coefficients $s_i$ for $i=1s,1c,2s,2c,3,4,7$ are associated with a CP-asymmetry in angular
coefficients: $J_i-\bar J_i$.

\item The coefficients $s_i$ for $i=5,6s,6c,8,9$ are associated with CP-averaged angular coefficients: $J_i+\bar J_i$.

\item The coefficients $s_i$ for $i=1s,1c,2s,2c,3,4,5,6s,6c$ are given by the imaginary part of amplitude interferences,
$s_i \sim \im (e^{i\phi} \bar A_X A_Y^*)$, and vanish in the absence of complex phases.
This is approximately true for $B_s\to V\ell\ell$ decays in the SM in
regions where strong phases are small, e.g. in the region $s\simeq 1-6$ GeV$^2$, and if the NP contribution has
the same weak phase as the SM.
The corresponding coefficients $J_i-\widetilde J_i$ do not vanish, in general.

\item The coefficient $s_7$ vanishes in the absence of phases in the amplitudes, while the combination $J_7-\bar J_7$
vanishes in the absence of CP violation in decay. Both are therefore very small in the SM, and also if the NP amplitudes
have approximately the same phase as the SM.

\item In the same conditions as above (no complex phases), the coefficients $(J_i+\bar J_i)_{i=8,9}$ vanish,
while $s_{8,9}$ do not.

\end{itemize}

It seems therefore that the most promising observables in this context are $s_{8,9}$,
which could be large and can be extracted from the time evolution of
\begin{equation}
J_8(t)-\widetilde J_8(t)\simeq -s_8\, e^{-\Gamma t} \sin(x\Gamma t)\ ,\qquad
J_9(t)-\widetilde J_9(t)\simeq -s_9\, e^{-\Gamma t} \sin(x\Gamma t)\ .
\end{equation}
The coefficients $s_8$ and $s_9$ have the following expressions (see Appendix~\ref{app2}):
\begin{eqnarray}
s_8&=& -\frac{1}{\sqrt{2}}\beta_\ell^2\,{\rm Re}\Big[e^{i\phi}(\bar A^{L}_{0}A^{L*}_{\perp}+\bar A^{R}_{0}A^{R*}_{\perp})+e^{-i\phi}(A^{L}_{0}\bar A^{L*}_{\perp}+A^{R}_{0}\bar A^{R*}_{\perp})\Big]\ ,\\[2mm]
s_9&=& \beta_\ell^2\,{\rm Re}\Big[e^{i\phi}(\bar A^{L}_{||}A^{L*}_{\perp}+\bar A^{R}_{||}A^{R*}_{\perp})+e^{-i\phi}(A^{L}_{||}\bar A^{L*}_{\perp}+A^{R}_{||}\bar A^{R*}_{\perp})\Big]\ .
\end{eqnarray}
We have checked by direct calculation that indeed the coefficients $s_i$ with $i\ne 8,9$ are tiny in the SM,
and that they do not get significant enhancement from NP contributions if new sources of CP violation are not large.

We emphasise that the measurement of the coefficients $s_{8,9}$ is challenging from the experimental point of view, since
the study of $J_i(t)-\widetilde J_i(t)$ requires
\emph{1)} flavour tagging of the original sample to separate $B$ and $\bar B$ at $t=0$,
\emph{2)} the use of appropriate foldings to extract the corresponding angular contributions,
identical to the ones used to extract $J_8$ and $J_9$ \cite{1304.6325,1308.1707}, and
\emph{3)} a time-dependent analysis to isolate the $\sin(x\Gamma t)$ coefficients.

\subsection{Symmetries of the distribution}
\label{sec:sym}

Having identified a few new observables accessible from the time-dependent angular distributions, it remains to be seen
if they are truly independent from the observables that can be extracted from angular  distributions
of flavour-specific decays.
The information that can be obtained from the angular distributions depends on the number of independent combinations
of interference terms $A_X A_Y^*$ in the angular coefficients.
A systematic formalism to determine which combinations can be accessed from the angular distributions alone is the
``symmetry formalism" developed in Refs.~\cite{1005.0571,1202.4266}.\footnote{See also Ref.~\cite{1502.00920} for an
application to S- and P-wave components in $B\to (K\pi)\mu\mu$.}

In the approximation of massless leptons, and neglecting scalar and tensor operators,
the angular distributions of flavour-specific decays contain a unitary symmetry,
given by the transformation \cite{1202.4266}:
\eq{n_i \equiv \binom{A_i^L}{\sigma_i A_i^{R*}} \rightarrow U n_i}
with $U$ an arbitrary unitary $2\times 2$ matrix, and $\{ \sigma_0, \sigma_\|, \sigma_\bot\} \equiv \{1,1,-1\}$.
Under this group of transformations, $J_i \to J_i$. This means that from flavour-specific decays,
only those combinations of terms $A_X A_Y^*$ that remain invariant under this transformation can be accessed.
This approach is useful to eliminate redundancies among observables built from the angular coefficients
$J_i$~\cite{1005.0571,1202.4266}.

We now identify the transformation properties of the coefficients $s_i$ --neglecting weak phases for simplicity.
We note that, under the unitary transformation:
\eqa{
\text{Re} \big[A_i^L A_j^{L*} \pm A_i^R A_j^{R*}\big] &\longrightarrow&
[1 - (1\mp \sigma_{ij})\lambda^2]\ \text{Re} \big[A_i^L A_j^{L*} \pm A_i^R A_j^{R*}\big]\nn\\[2mm]
&& + (1\mp \sigma_{ij})\ \text{Re} \big[\sigma_i \eta A_i^R A_j^L + \sigma_j \eta A_i^L A_j^R\big]\ ,\\[4mm]
\text{Im} \big[A_i^L A_j^{L*} \pm A_i^R A_j^{R*}\big] &\longrightarrow&
[1 - (1\pm \sigma_{ij})\lambda^2]\ \text{Im} \big[A_i^L A_j^{L*} \pm A_i^R A_j^{R*}\big]\nn\\[2mm]
&& - (1\pm \sigma_{ij})\ \text{Im} \big[\sigma_i \eta A_i^R A_j^L - \sigma_j \eta A_i^L A_j^R\big]\ ,
}
where $\sigma_{ij}=\sigma_i\sigma_j$, $\lambda^2 \equiv 1-|U_{11}|^2$, $\eta \equiv U_{11} U_{12}^*$
and $i,j = 0,\|,\bot$. Non-trivial transformations involve only
$\text{Re} \big[A_i^L A_j^{L*} \pm A_i^R A_j^{R*}\big]$ with $\sigma_{ij}=\mp 1$, or else
$\text{Im} \big[A_i^L A_j^{L*} \pm A_i^R A_j^{R*}\big]$ with $\sigma_{ij}=\pm 1$.
From the explicit expressions given in Appendix~\ref{app2}, we see that
(neglecting lepton mass terms and weak phases in the amplitudes):
\eqa{
s_{1s,2s} &\sim& \sin\phi \cdot \re \big[ A_\|^L A_\|^{L*} + A_\|^R A_\|^{R*}  \big] - (\| \to \bot)\ ,\\[2mm]
s_{1c,2c} &\sim& \sin\phi \cdot \re \big[ A_0^L A_0^{L*} + A_0^R A_0^{R*}  \big]\ ,\\[2mm]
s_{3} &\sim& \sin\phi \cdot \re \big[ A_\|^L A_\|^{L*} + A_\|^R A_\|^{R*}  \big] + (\| \to \bot)\ ,\\[2mm]
s_{4} &\sim& \sin\phi \cdot \re \big[ A_0^L A_\|^{L*} + A_0^R A_\|^{R*}  \big]\ ,\\[2mm]
s_{5} &\sim& \cos\phi \cdot \im \big[ A_0^L A_\bot^{L*} - A_0^R A_\bot^{R*}  \big]\ ,\\[2mm]
s_{6s} &\sim& \cos\phi \cdot \im \big[ A_\|^L A_\bot^{L*} - A_\|^R A_\bot^{R*}  \big]\ ,\\[2mm]
s_{7} &\sim& \sin\phi \cdot \im \big[ A_0^L A_\|^{L*} - A_0^R A_\|^{R*}  \big]\ ,\\[2mm]
s_{8} &\sim& \cos\phi \cdot \re \big[ A_0^L A_\bot^{L*} + A_0^R A_\bot^{R*}  \big]\ ,\\[2mm]
s_{9} &\sim& \cos\phi \cdot \re \big[ A_\|^L A_\bot^{L*} + A_\|^R A_\bot^{R*}  \big]\ .
}
Therefore the only coefficients $s_i$ that (in this approximation) do not remain
invariant are $s_5$, $s_{6s}$, $s_8$ and $s_9$, which contain additional information not accessible from the usual
angular distributions of flavour-specific decays such as $B_d\to K^*(\to K^+\pi^-)\ell\ell$. Among these coefficients,
we have seen that $s_{8,9}$ are particularly promising; now we see that they are independent from the coefficients $J_i$.

\section{Observables}
\label{sec:3}

The expressions in Eqs.~(\ref{eq:J+Jt}),~(\ref{eq:J-Jt}) for the coefficients of the time-dependent distributions,
show that additional structures arise in the presence of neutral-meson mixing. In this context, two different
quantities might be considered: time-integrated observables, or observables related to the time dependence.
In this section we discuss the two possibilities.

\subsection{Time-integrated observables}
\label{sec:t-int}

As discussed in Refs.~\cite{Harrison:1998yr,1111.4882}, time integration should be performed differently in the context of
hadronic machines and $B$-factories. The time-dependent expressions in Eqs.~(\ref{eq:J+Jt}) and (\ref{eq:J-Jt})
are written in the case of tagging at a
hadronic machine, assuming that the two $b$-quarks have been produced incoherently, with $t\in [0 , \infty)$.
In the case of a coherent $B \bar{B}$ pair produced at a $B$-factory, one must replace $\exp(-\Gamma t)$
by $\exp(-\Gamma |t|)$ and integrate over $t\in (-\infty,\infty)$~\cite{Harrison:1998yr}.
Interestingly, the integrated versions of CP-violating interference terms are different in both settings,
and the measurement at hadronic machines involves an additional term compared to the $B$-factory case:
\allowdisplaybreaks{
\begin{eqnarray}
\label{eq:<J+Jt>Had}
\langle J_i + \widetilde J_i\rangle_{\rm Hadronic}
 &=& \frac{1}{\Gamma} \left[\frac{1}{1-y^2} \times(J_i+\widetilde J_i)-\frac{y}{1-y^2}\times h_i\right]\ ,\\
\label{eq:<J-Jt>Had}
\langle J_i - \widetilde J_i\rangle_{\rm Hadronic}
 &=& \frac{1}{\Gamma}\Bigg[\frac{1}{1+x^2} \times (J_i-\widetilde J_i) - \frac{x}{1+x^2} \times s_i \Bigg]\ ,\\
 \label{eq:<J+Jt>Bfac}
\langle J_i + \widetilde J_i\rangle_{\rm B-factory}
 &=& \frac{2}{\Gamma}\frac{1}{1-y^2}[J_i+\widetilde J_i]\ ,\\
  \label{eq:<J-Jt>Bfac}
 \langle  J_i - \widetilde J_i\rangle_{\rm B-factory}
 &=& \frac{2}{\Gamma}\frac{1}{1+x^2} [J_i-\widetilde J_i]\ .
\end{eqnarray}
}
 
Making contact with experimental measurements requires to consider the total time-integrated decay rate.
The time-dependent rate is given by
\begin{equation}
\frac{d\Gamma}{dq^2}=\int dt\ \left[\frac{3}{4}\big(2J_{1s}(t)+J_{1c}(t)\big)
-\frac{1}{4}\big(2J_{2s}(t)+J_{2c}(t)\big)\right]\ ,
\end{equation}
which after time-integration becomes 
\begin{eqnarray}
\left\langle\frac{d\Gamma}{dq^2}\right\rangle &=&  \frac{1}{\Gamma(1-y^2)} \langle{\cal I}\rangle\ ,\\[2mm]
\langle{\cal I}\rangle _{\rm Hadronic}
&=&\frac{3}{4}\bigg[2(J_{1s}+\bar{J}_{1s}-y\,h_{1s})+(J_{1c}+\bar{J}_{1c}-y\,h_{1c}) \bigg] \nn\\
&&-\frac{1}{4}\bigg[2(J_{2s}+\bar{J}_{2s}-y\,h_{2s})+(J_{2c}+\bar{J}_{2c}-y\,h_{2c}) \bigg]\ ,\\[2mm]
\langle{\cal I}\rangle _{\rm B-factory} &=&\langle{\cal I}\rangle_{\rm Hadronic}(h=0)\ ,
\end{eqnarray}
where ${\cal I}$ is the usual normalisation considered in analyses of the angular coefficients.
The normalised time-integrated angular coefficients at hadronic machines or $B$-factories are therefore:
\begin{eqnarray}
\langle \Sigma_i\rangle_{\rm Hadronic}&\equiv&
\frac{\langle J_i + \widetilde J_i\rangle_{\rm Hadronic}}{\langle d\Gamma/dq^2\rangle_{\rm Hadronic}}
=\frac{(J_i+\widetilde J_i)-y\times h_i}{\langle{\cal I}\rangle_{\rm Hadronic}}\ ,\\
\langle \Sigma_i\rangle_{\rm B-factory}&\equiv&
\frac{\langle J_i + \widetilde J_i\rangle_{\rm B-factory}}{\langle d\Gamma/dq^2\rangle_{\rm B-factory}}
=\langle \Sigma_i\rangle_{\rm Hadronic}(h=0)\ ,\\
\langle \Delta_i\rangle_{\rm Hadronic}&\equiv&
\frac{\langle J_i - \widetilde J_i\rangle_{\rm Hadronic}}{\langle d\Gamma/dq^2\rangle_{\rm Hadronic}}
=\frac{1-y^2}{1+x^2}\times \frac{(J_i-\widetilde J_i)-x\times s_i}{\langle{\cal I}\rangle_{\rm Hadronic}}\ ,\\
\langle \Delta_i\rangle_{\rm B-factory}&\equiv&
\frac{\langle J_i - \widetilde J_i\rangle_{\rm B-factory}}{\langle d\Gamma/dq^2\rangle_{\rm B-factory}}
=\langle \Delta_i\rangle_{\rm Hadronic}(h=s=0)\ .
\end{eqnarray}

We see that the interpretation of the time-integrated measurements $\langle \Sigma_i\rangle$ from
$d\Gamma(B\to f_{CP}\ell\ell)+d\Gamma(\bar{B}\to f_{CP}\ell\ell)$
is straightforward in terms of the angular coefficients at $t=0$. Even in the $B_s$ case, the smallness of $y$
means that $h_i$ will have only a very limited impact on the discussion. 
The time-integrated terms $\langle \Delta_i\rangle$  from $d\Gamma(B\to f_{CP}\ell\ell)-d\Gamma(\bar{B}\to f_{CP}\ell\ell)$
are subject to two different effects, in particular for $B_s$ where $x$ is large:
\begin{itemize}

\item[(a)] they receive contributions proportional to $x$ and $y$ with a different combination of interference
terms (in the case of a measurement at a hadronic machine),

\item[(b)] they are suppressed (in all experimental set-ups) by a factor $(1-y^2)/(1+x^2)$. 

\end{itemize}

The discussion above applies in particular to the measurement of $B_s\to \phi(\to K^+K^-)\ell\ell$ as performed at
LHCb~\cite{1305.2168}. Since this is not a self-tagging mode, and assuming that there is an equal production of
$B_s$ and $\bar{B}_s$, what is measured is
$d\Gamma(B_s\to \phi(\to K^+K^-)\ell\ell)+d\Gamma(\bar{B}_s\to \phi(\to K^+K^-)\ell\ell)$, so these measurements
have access to the following combinations:
\eqa{
\langle J_i + \bar J_i\rangle_{\rm Hadronic} \quad &\text{for}\ &  i=1s,1c,2s,2c,3,4,7,\nn\\
\langle J_i - \bar J_i\rangle_{\rm Hadronic} \quad &\text{for}\ &  i=5,6s,6c,8,9.
}
The time-integrated observables $\langle \Sigma_6\rangle_{\rm Hadronic}$ and $\langle \Sigma_9\rangle_{\rm Hadronic}$
have already been measured (under the name of $A_6$ and $A_9$) in Ref.~\cite{1305.2168}, and are indeed measuring
CP-violation.
In the context of the extraction of $s_8$ and $s_9$ at hadronic machines, one expects them to
dominate $\langle \Delta_{8,9}\rangle_{\rm Hadronic}$, especially in the case of $B_s$ decays where $x$ enhances
their contribution with respect to the $(J_i - \widetilde J_i)$ term. However, they are overall suppressed by
a factor $\sim 1/x$, which in the case of $B_s$ decays is quite effective ($1/x\sim 0.04$).
In addition, the necessity to perform initial flavour tagging makes these measurements very difficult
at hadronic machines.

We see therefore that $\av{\Sigma_i}$ contain essentially the same information as $(J_i+\widetilde J_i)$, whereas
$\av{\Delta_i}$ have a potentially richer interpretation, but are suppressed and thus probably difficult to 
extract experimentally. In the following section we will see that time-dependent observables
do lead to more interesting opportunities.

\subsection{Time-dependent ``optimised" observables with tagging}
\label{sec:Q8Q9}

From the discussion in Section~\ref{sec:2.4} is clear that a full tagged time-dependent angular analysis
of the decay $B\to V[\to (M_1M_2)_{CP}]\ell\ell$ provides a measurement of the angular observables $J_i(s)$,
$\bar J_i(s)$, $s_i(s)$ and $h_i(s)$,
i.e. Eqs.~(\ref{G+Gb}),~(\ref{GmGb}),~(\ref{eq:J+Jt}),~(\ref{eq:J-Jt}).
We have also seen that the coefficients $h_i$ are the $\sinh(y\Gamma t)$ coefficient of
$J_i(t)+\widetilde J_i(t)$, whose effect remains negligible for $t\lesssim 10\,\tau_{B_q}$, constituting
a rather difficult measurement.

From the theoretical point of view, these observables are quadratic in hadronic form factors (see Section~\ref{sec:SM}).
For instance,
\eq{
s_i(s) \sim A_X A_Y^* \sim F_X^{B\to V}(s)\cdot F_Y^{B\to V}(s)\ ,
}
and similarly for $J_i(s),\bar J_i(s), h_i(s)$. Here $F_{X,Y}$ represent (schematically) hadronic $B\to V$ form factors
related to the amplitudes $A_{X,Y}$. These form factors constitute a major source of uncertainty
in the theoretical predictions for the observables. This problem is usually tamed by
defining a class of special observables with reduced sensitivity to form-factor uncertainties 
\cite{0502060,1006.5013,1105.0376,1106.3283,1202.4266,1207.2753,1303.5794}.
These ``optimised" observables can be constructed systematically, both in the region of large recoil of
the vector meson ($s\ll m_B^2$) \cite{1202.4266} and at low recoil ($s \sim m_B^2$)
\cite{1006.5013,1303.5794}, where the use of effective field theories
(SCET \cite{9812358,0008255} and HQET \cite{0404250,1101.5118} respectively)
ensures a complete cancellation of form factors at the leading order in the respective expansions\footnote{
In the following and throughout the paper we use the term \emph{``large-recoil limit"} to denote the
following approximation valid in the region $s\ll m_B^2$: leading order in $\alpha_s$ and leading
power in the SCET expansion. This is, of course, a slight abuse of language.}.

In the following, we focus on the large-recoil region for definiteness. We consider the following
optimised versions of the observables $s_{8,9}$:
\eqa{
\label{Q8m}
Q_8^- &=& \frac{s_8}{\sqrt{-2(J_{2c}+\widetilde J_{2c}) [2(J_{2s}+\widetilde J_{2s})-(J_3+\widetilde J_3)]}}\ ,\\[2mm]
\label{Q9}
Q_9 &=& \frac{s_9}{2 (J_{2s}+\widetilde J_{2s})}\ .
}
There are other possible normalizations for $s_8$ that are also optimised at large recoil:
\eqa{
Q_8^+ &=& \frac{s_8}{\sqrt{-2(J_{2c}+\widetilde J_{2c}) [2(J_{2s}+\widetilde J_{2s})+(J_3+\widetilde J_3)]}}\ ,\\[2mm]
Q_8^0 &=& \frac{s_8}{\sqrt{-2(J_{2c}+\widetilde J_{2c}) [2(J_{2s}+\widetilde J_{2s})]}}\ .
}
The observable $Q_8^+$ has the particularity of being also optimised at low recoil.
However, we find that both $Q_8^+$ and $Q_8^0$ are slightly less sensitive to NP than $Q_8^-$.
While it might be worthwhile to study these observables further, we will focus here on $Q_8^-$
for illustration, noting that its properties do not differ much from those of $Q_8^+$, $Q_8^0$.

Concerning $Q_9$, other possible normalizations involve $J_{6s}$ or $J_9$, both of which lead to
observables that are optimised also at low recoil. We do not consider these possibilities any further as the
denominators contain zeroes within the kinematical region of interest. 

It is useful to consider these observables in the large-recoil limit (see e.g. Refs.~\cite{1104.3342,1202.4266}),
where the expressions simplify considerably, the cancellation of form factors is exact, and the dependence on
the Wilson coefficients is apparent. We find:
\eqa{
Q_8^- &=& \left[ \frac{\C7^+ (2 \C7^- + \C9^-)}{2|\C7^-| \sqrt{(\C{10}^-)^2+(2 \C7^- + \C9^-)^2}}\right.\nn\\[2mm]
&& \left. + \frac{(\C{10}^- \C{10}^+ \C7^- + (2 \C7^- + \C9^-) (-\C7^+ \C9^- + \C7^- \C9^+))}
{4\C7^- |\C7^-|\sqrt{(\C{10}^-)^2+(2 \C7^- + \C9^-)^2}} \frac{s}{m_B^2} + \cdots
\right]\cos{\tilde \phi_q}\\[4mm]
Q_9 &=& -\left[
\frac{2 \C7^- \C7^+ }{(\C7^-)^2 + (\C7^+)^2}
+ \frac{[(\C7^-)^2 - (\C7^+)^2] (\C7^- \C9^+ - \C7^+ \C9^-)}{[(\C7^-)^2 + (\C7^+)^2]^2}
\frac{s}{m_B^2} + \cdots
\right] \cos{\tilde \phi_q}
}
where we have assumed real $\C{i}$, and used the notation $\C{i}^\pm = \C{i} \pm \C{i'}$.
In the case of the $b\to s$ processes at hand, we have $\tilde\phi_q \equiv \phi_q - 2\beta_s$, with $\phi_q$
the mixing angle in the $B_q$ system.
We note that if $\C{i'}=0$ (that is, $\C{i}^+=\C{i}^-$), on has $Q_9=-\cos\tilde\phi_q$,
so that the value of $(Q_9+\cos\tilde\phi_q)$ is a measurement of right-handed currents.

In the following section we give Standard Model predictions for these observables and study briefly their
sensitivity to New Physics.

\section{Numerical Analysis}
\label{sec:4}

\subsection{Standard Model}
\label{sec:SM}

The systematic formalism to $B\to V\ell\ell$ decays at large hadronic recoil to NLO in QCD-factorisation has been
presented in Ref.~\cite{0106067} and is by now quite standard. In our analysis we follow closely the procedure
of Refs.~\cite{1303.5794,1407.8526} to which we refer the reader for further details.
The different transversity amplitudes can be written as:
\eqa{
\label{Abot}
A_\bot^{L,R}(s) &=&
{\cal N}_\bot \ \bigg\{
\Big[ (\C9^+ + Y_t(s)(1+\eta_{\sss\rm PC}^{(1)}) + \lambda_{ut} Y_u(s)) \mp \C{10}^+ \Big] \frac{V(s)}{M+m}
+ \frac{2m_b}{s} {\cal T}_1^+ \bigg\}\ ,\\[2mm]
A_\|^{L,R}(s) &=&
{\cal N}_\| \ \bigg\{
\Big[ (\C9^- + Y_t(s)(1+\eta_{\sss\rm PC}^{(2)}) + \lambda_{ut} Y_u(s)) \mp \C{10}^- \Big] \frac{A_1(s)}{M+m}
+ \frac{2m_b}{s} {\cal T}_2^- \bigg\}\ ,\\[2mm]
A_0^{L,R}(s) &=&
{\cal N}_0 \ \bigg\{
\Big[ (\C9^- + Y_t(s)(1+\eta_{\sss\rm PC}^{(3)}) + \lambda_{ut} Y_u(s)) \mp \C{10}^- \Big] \frac{A_{12}(s)}{M+m}\nn\\
&& \hspace{3.5cm} + \frac{2m_b}{s} \Big[ (M^2+3 m^2-s){\cal T}_2^- - \frac{\lambda}{M^2-m^2} {\cal T}_3^-\Big] \bigg\}\ ,
\label{A0}\\
A_t(s) &=& \frac{{\cal N}_\bot}{\sqrt{2s}} \ \Big[  2\C{10}^- + \frac{s}{2m_\ell} \C{P}^- \Big] A_0(s)\ ,
\qquad A_S(s) = - {\cal N}_\bot\ \C{S}^-\,A_0(s)\ ,
}
where $M=m_{B_q}$ and $m=m_V$, and:

\begin{table}
\begin{center}
\small
\begin{tabular}{cccccccccc}
\hline
$\!\C1(\mu_b)\!$ &   $\!\C2(\mu_b)\!$ &  $\!\C3(\mu_b)\!$ &  $\!\C4(\mu_b)\!$
& $\!\C5(\mu_b)\!$ & $\!\C6(\mu_b)\!$ & $\!\C7^{\rm eff}(\mu_b)\!$ & $\!\C8^{\rm eff}(\mu_b)\!$
& $\!\C9(\mu_b)\!$ & $\! \C{10}(\mu_b)\!$ \\
\hline
-0.2632 & 1.0111 & -0.0055 & -0.0806 & 0.0004 &
0.0009 &  -0.2923 & -0.1663 & 4.0749 & -4.3085\\
\hline
\end{tabular}
\caption{Wilson coefficients in the Standard Model at NNLO at the scale $\mu_b=4.8\,{\rm GeV}$.}
\label{tab:Ci}
\end{center}
\end{table}

\begin{itemize}

\item The normalizations are given by
\eq{
{\cal N}_\bot=\sqrt{2\lambda} N\ ,\quad {\cal N}_\|=\sqrt{2} (M^2-m^2) N\ ,\quad {\cal N}_0 = -N/(2 m\sqrt{s})\ ,
}
with $\lambda=M^4+m^4+s^2-2(M^2m^2+M^2 s+m^2 s)$, $\beta_\ell=\sqrt{1-4m_\ell^2/s}$, and
\eq{
N(B)=V^*_{tb}V_{ts} \bigg[ \frac{G_F^2 \alpha^2\,s\,\lambda^{1/2} \beta_\ell}{3\cdot 2^{10} \pi^5 M^3} \bigg]^{1/2}\ ,
\quad
N(\bar B)=V_{tb}V^*_{ts} \bigg[ \frac{G_F^2 \alpha^2\,s\,\lambda^{1/2} \beta_\ell}{3\cdot 2^{10} \pi^5 M^3} \bigg]^{1/2}\ .
}

\item $Y_t(s)$ and $Y_u(s)$ are the 1-loop contributions from 4-quark operators to the photon penguin with the structure
$\bar s \gamma_\mu b$, sometimes combined with $\C9$ into $\C{9\text{eff}}(s)$. $Y_t(s)$ denotes the contribution proportional to $V_{tb} V_{ts}^*$, and can be found in
Eq.(10) of Ref.~\cite{0106067}. $Y_u(s)$ denotes the CKM-suppressed contribution, which is multiplied by the prefactor
$\lambda_{ut}(B)= V_{ub} V_{us}^*/ V_{tb} V_{ts}^*$ or $\lambda_{ut}(\bar B)= V_{ub}^* V_{us}/V_{tb}^* V_{ts}$.
This function can be found in Eq.~(A.3) of Ref.~\cite{0412400}.

\item The functions ${\cal T}_i$ encode contributions from dipole operators $\C{7^{(\prime)}}$,
and the rest of the hadronic contributions not contained in $Y_t,Y_u$:
\eqa{
{\cal T}_1^+(s) &=& \C{7\text{eff}}^+\,T_1(s) + {\cal T}_\bot (s) (1+\eta_{\sss\rm PC}^{(1)})\label{Tau1}\\
{\cal T}_2^-(s) &=& \C{7\text{eff}}^-\,T_2(s) + \frac{M^2-s}{M^2} {\cal T}_\bot (s) (1+\eta_{\sss\rm PC}^{(2)})\\
{\cal T}_3^-(s) &=& \C{7\text{eff}}^-\,T_3(s) + {\cal T}_\bot (s) + {\cal T}_\| (s) (1+\eta_{\sss\rm PC}^{(3)})\label{Tau3}
}
The quantities ${\cal T}_{\bot,\|}$ represent factorizable and non-factorizable hadronic contributions in
QCD-factorisation and can be extracted from the formulae in Section 2 of Ref.~\cite{0106067}. They depend
on distribution amplitudes and on two ``soft" form factors $\xi_\bot(s)$, $\xi_\|(s)$.
The ``effective" coefficients $\C{7\text{eff}}^\pm$ include contributions from 4-quark operators with $b$ and $s$-quark
loops with the structure $\bar s [/\!\!\!q,\gamma_\mu] b$ (see e.g. Refs.~\cite{9711280,1411.7677}).

\item The parameters $\eta_{\sss\rm PC}^{(i)}$ in Eqs.~(\ref{Abot})-(\ref{A0})~and~(\ref{Tau1})-(\ref{Tau3})
parametrize non-factorizable $\op(\Lambda/m_b)$ power corrections absent in the current QCD-factorisation calculation.
Following Ref.~\cite{1407.8526}, we write
\eq{
\eta_{\sss\rm PC}^{(i)} = r_i^a e^{i\phi_i^a} + r_i^b e^{i\phi_i^b} (s/M^2) + r_i^c e^{i\phi_i^c} (s/M^2)^2
}
and take $r_i^{a,b,c}=0$ as our central value, varying the parameters $r_i,\phi_i$ within the ranges
$r_i^{a,b,c}\in [0,0.1]$ and $\phi_i^{a,b,c}\in [-\pi,\pi]$ in the error analysis. This corresponds to a
contribution from non-factorizable $\op(\Lambda/m_b)$ corrections of $\op(10\%)$ with an arbitrary phase.

\item The functions $V(s), A_0(s), A_1(s), A_2(s), T_1(s), T_2(s), T_3(s)$ represent the seven independent $B\to V$
QCD form factors (see e.g. Refs.~\cite{9812358,0008255}), with the combination
\eq{
A_{12}(s)=(M^2-m^2-s)(M+m)^2 A_1(s)-\lambda A_{2}(s)
}
entering $A_0^{L,R}$. Following Ref.~\cite{1303.5794}, we use the large-recoil symmetry relations \cite{9812358,0008255}
to express these form factors in terms of $\xi_\bot(s)$ and $\xi_\|(s)$, defined in the ``scheme 1" of
Ref.~\cite{1407.8526}, including factorizable power corrections. At a second stage, these are themselves expressed
in terms of $V(s), A_1(s), A_2(s)$, which are taken from the light-cone sum rule calculation of
Ref.~\cite{0412079}, both for $B\to K^*$ and $B_s\to \phi$ transitions.

\item The Wilson coefficients $\C{i}^\pm$ are defined as: $\C{i}^\pm = \C{i} \pm \C{i'}$. The Standard Model values
for these coefficients are collected in Table~\ref{tab:Ci}, computed at a renormalisation scale $\mu_b=4.8\,{\rm GeV}$.
As has become customary in analyses of $B\to V\ell\ell$
decays~\cite{1307.5683,1308.1501,1310.2478,1310.3887,1408.1627,1408.4097,1410.4545,1411.3161},
we use NNLO Wilson coefficients, keeping in mind that
the NNLO scheme and scale ambiguity can only be eliminated by including the (currently unknown) NNLO matrix elements.
In this context, in the error analysis we consider a variation of the renormalisation scale $\mu\in [\mu_b/2,2\mu_b]$.
In addition, we have $\C{7'}^{\sss\rm SM}=(m_s/m_b)\,\C{7}^{\sss\rm SM}$.

\end{itemize}

Following this procedure, we compute central values and errors (by means of a flat scan over all parameters) for the
observables $Q_8^-$ and $Q_9$ in the SM. We compute these observables differentially in $s$, keeping in mind that
a proper comparison with data would require an integration over bin ranges (see for example the discussion
in Ref.~\cite{1207.2753}).

\begin{figure}
\includegraphics[width=7.5cm,height=4.9cm]{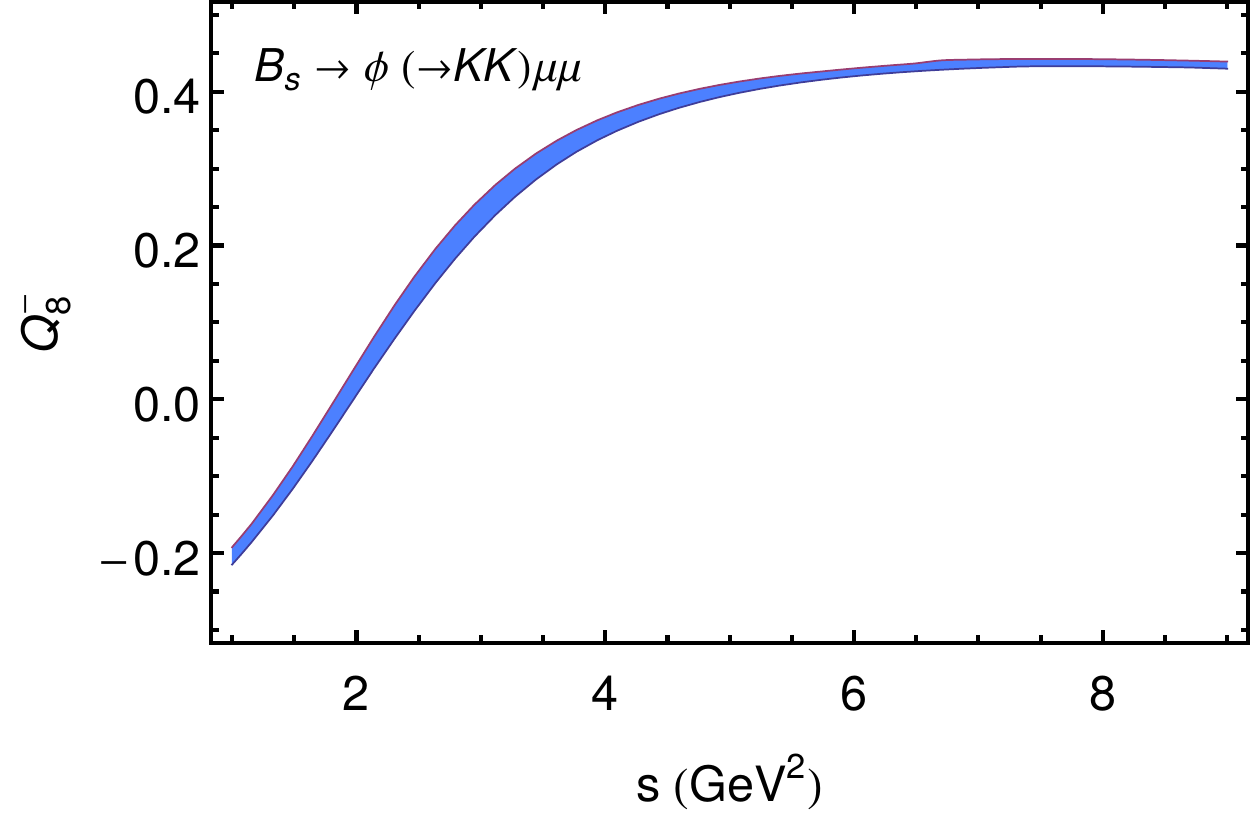}\hspace{4mm}
\includegraphics[width=7.5cm,height=5cm]{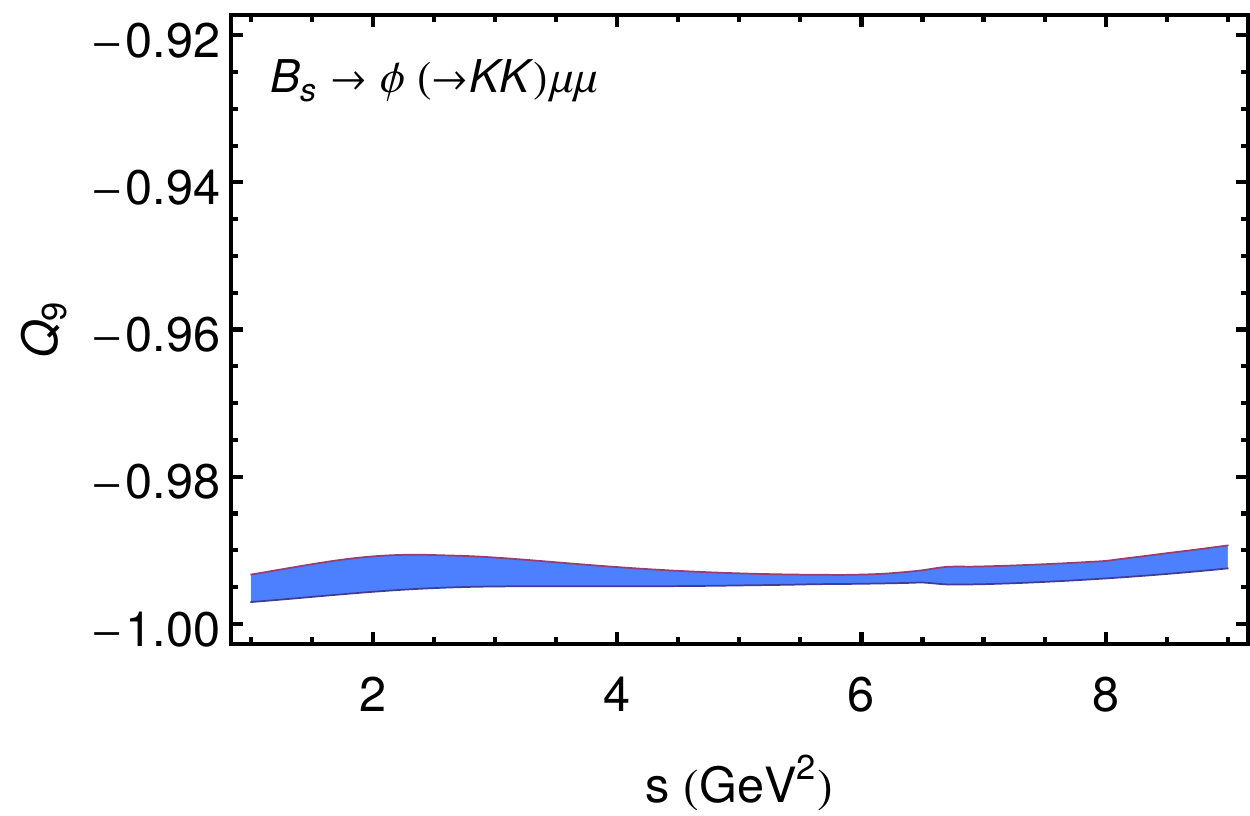}\\[3mm]
\includegraphics[width=7.5cm,height=4.9cm]{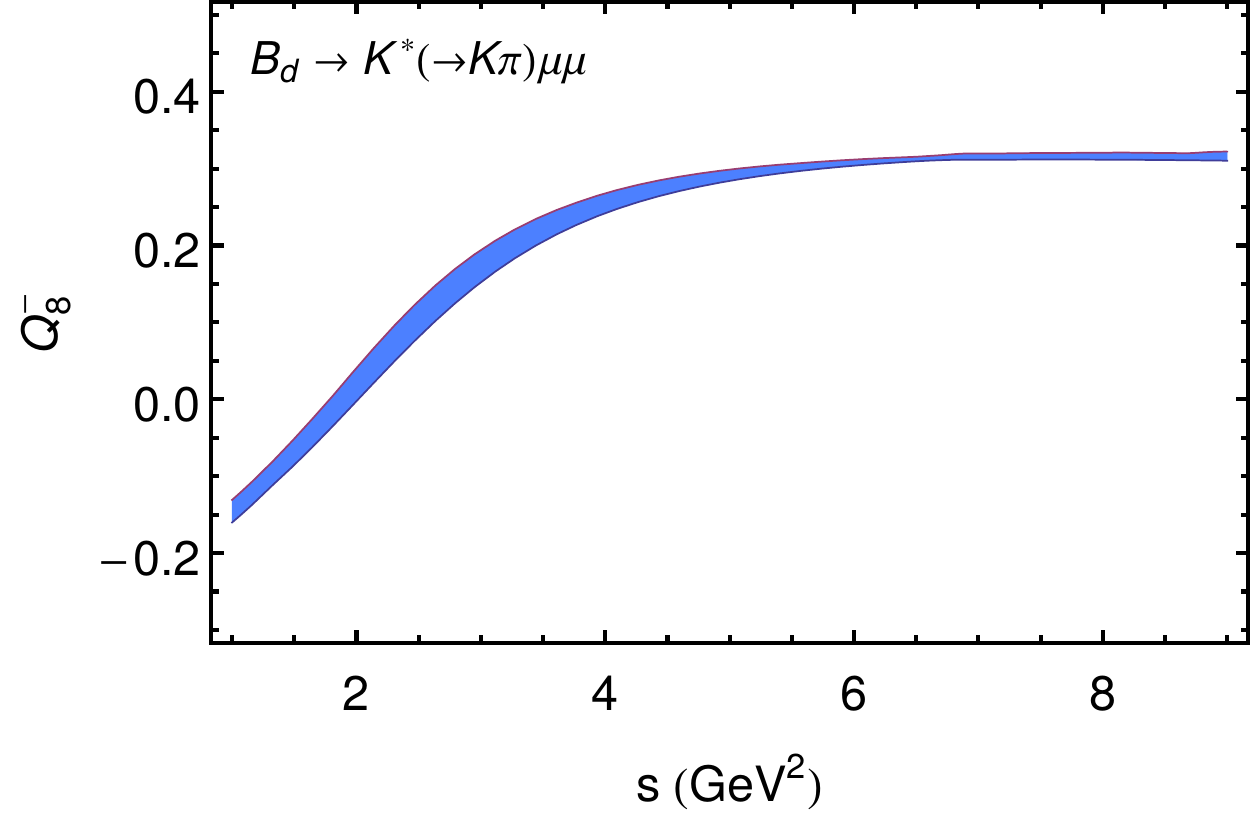}\hspace{4mm}
\includegraphics[width=7.5cm,height=5cm]{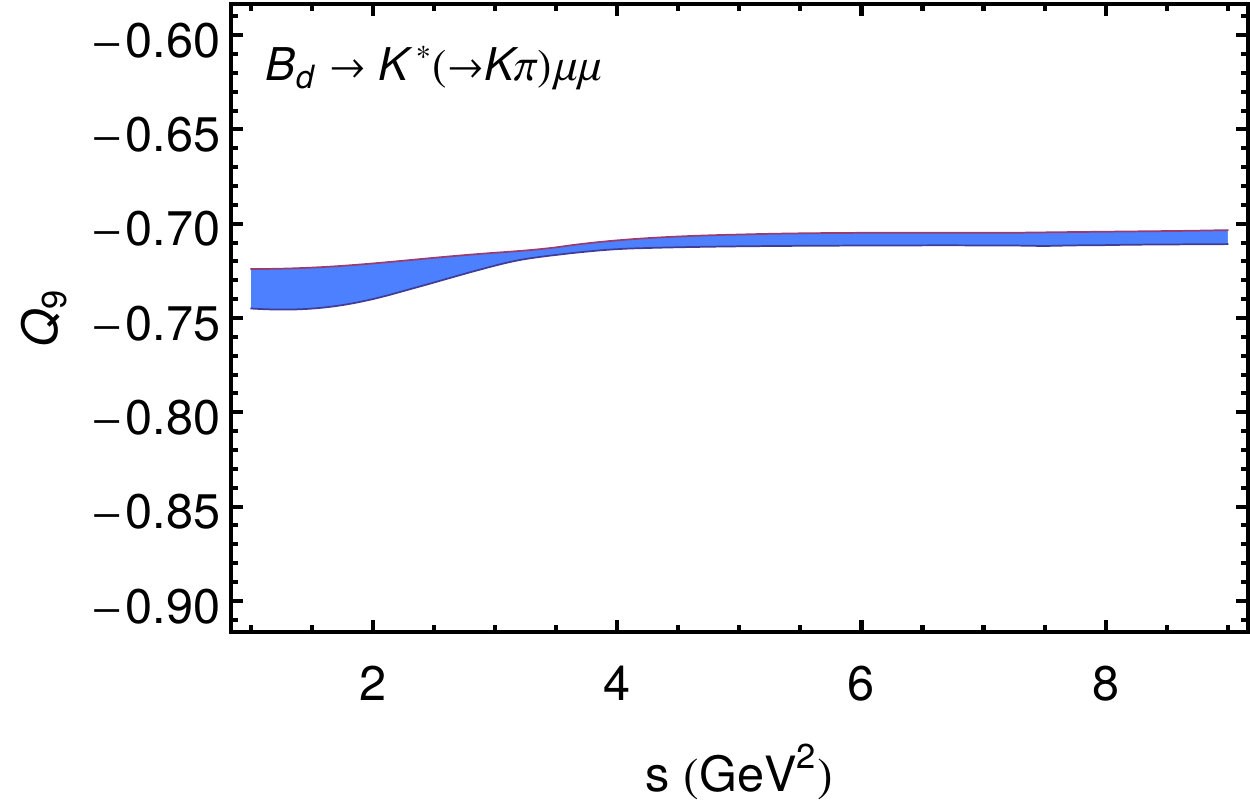}\\[-8mm]
\caption{SM prediction for the observables $Q_8^-$ and $Q_9$ in the case of $B_s(t)\to \phi(\to K^+K^-)\mu^+\mu^-$
(upper row) and $B_d(t)\to K^*(\to K_S\pi)\mu^+\mu^-$ (lower row),
in the large-recoil region, including error estimates from all sources. See the text for details.}
\label{figQ8Q9}
\end{figure}

Our SM results for the observables $Q_8^-$ and $Q_9$ in the low-$s$ region are shown in Fig.~\ref{figQ8Q9}.
We show both cases: $B_s(t)\to \phi(\to KK)\mu^+\mu^-$ and
$B_d(t)\to K^*(\to K_S\pi^0)\mu^+\mu^-$, noting that the results are very similar.
We see that indeed the observable $Q_9\simeq -\cos\tilde\phi_q$ in the whole region (with $\tilde\phi_d \simeq -2\beta$
and $\tilde\phi_s =0$ in the SM),
while $Q_8^-$ features a distinctive shape with a zero at $s_0\simeq 2$ GeV$^2$. 
This is located at the same position as the zero of $s_8$, which can be expressed solely in terms of
Wilson coefficients taking the large-recoil limit,
\eq{
\frac{s_0}{m_B^2} \simeq  \frac{-2 \C7^+ (2\C7^-+\C9^-)}{\C{10}^-\C{10}^++(2\C7^-+\C9^-)\C9^+}
\stackrel{\rm SM}{\simeq} \frac{-2 \C7 (2\C7+\C9)}{\C{10}^2+(2\C7+\C9)\C9}\ .
}
The position of this zero measures a different ratio of Wilson coefficients compared to the zero of
other observables, such as $A_\text{FB}$ or $P_2$~\cite{1202.4266}.
We stress that the bands in Fig.~\ref{figQ8Q9} include all sources
of error including parametric and form-factor uncertainties, as well as our estimates of power corrections,
exhibiting the theoretical accuracy for these observables in the Standard Model.

\subsection{New Physics}

We now study the sensitivity of the observables $Q_8^-$ and $Q_9$ to different models of New Physics.
We start with a general scan of (real) New Physics contributions to Wilson coefficients compatible with
all current constraints from rare $B$-decays, in order to assess the NP reach of the new observables.
For that purpose, we write
\eq{\C{i}=\C{i}^{\sss\rm SM} + \C{i}^{\sss\rm NP}\ ,}
and consider the $3\,\sigma$ ranges for the NP contributions $\C{i}^{\sss\rm NP}$ (at the scale $\mu_b=\op(m_b)$)
that were obtained in the global fit to $b\to s\gamma$ and $b\to s \ell\ell$ data of Ref.~\cite{1307.5683}:
\begin{align}
\C{7}^{\sss\rm NP} \in (-0.08,0.03)\ , &&
\C{9}^{\sss\rm NP} \in (-2.1,-0.2)\ , &&
\C{10}^{\sss\rm NP} \in (-2.0,3.0)\ , \nn\\
\C{7'}^{\sss\rm NP} \in (-0.14,0.10)\ , &&
\C{9'}^{\sss\rm NP} \in (-1.2,1.8)\ , &&
\C{10'}^{\sss\rm NP} \in (-1.4,1.2)\ .
\end{align}
The result of this scan is shown in Fig.~\ref{figNPscan}. We consider separately three scenarios:

\begin{figure}
\includegraphics[width=7.5cm,height=5cm]{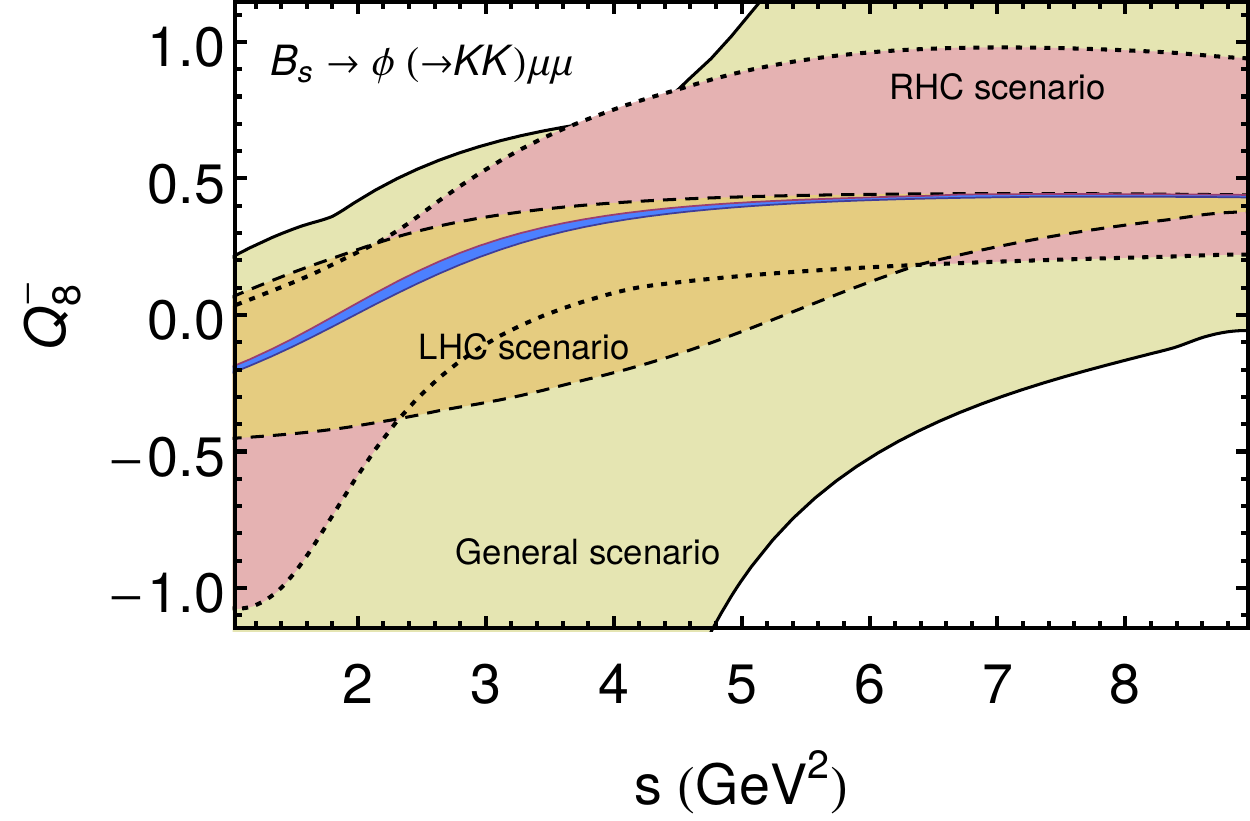}\hspace{4mm}
\includegraphics[width=7.5cm,height=5cm]{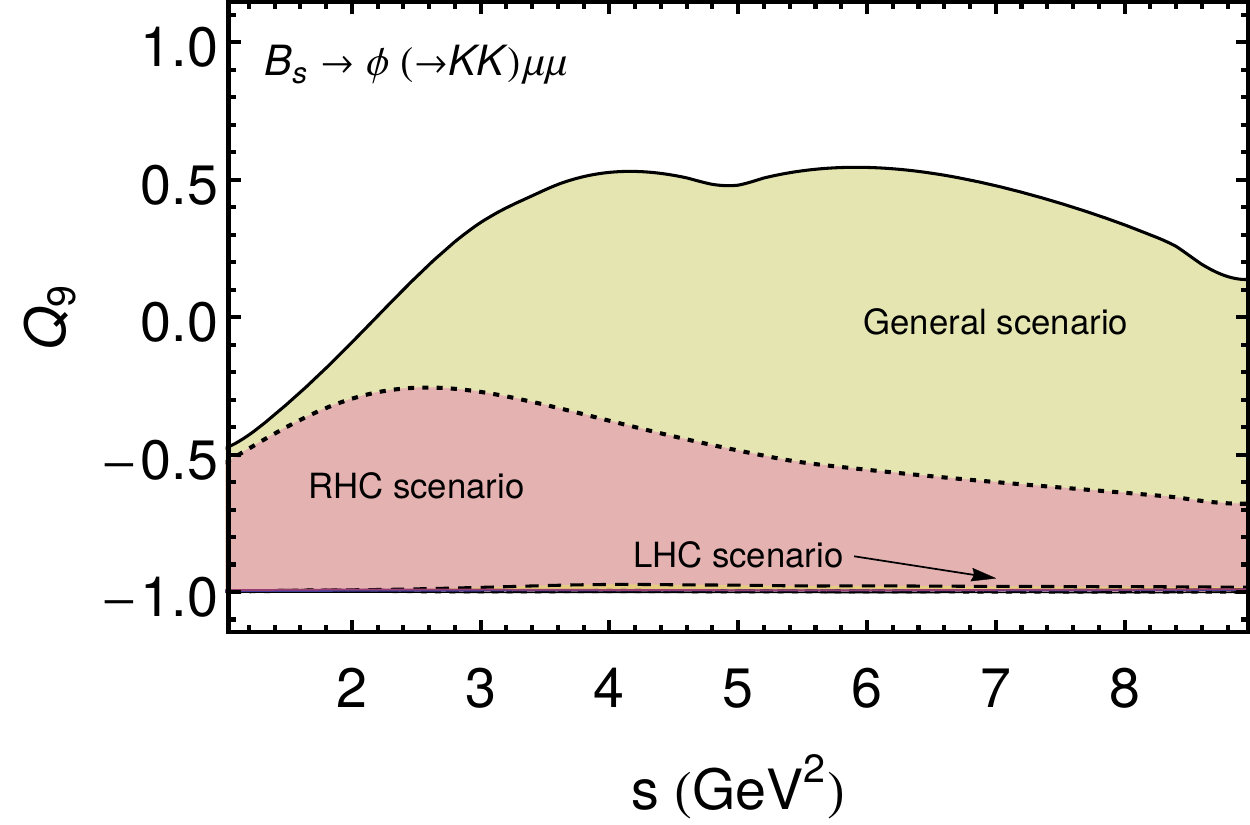}\\[4mm]
\includegraphics[width=7.5cm,height=5cm]{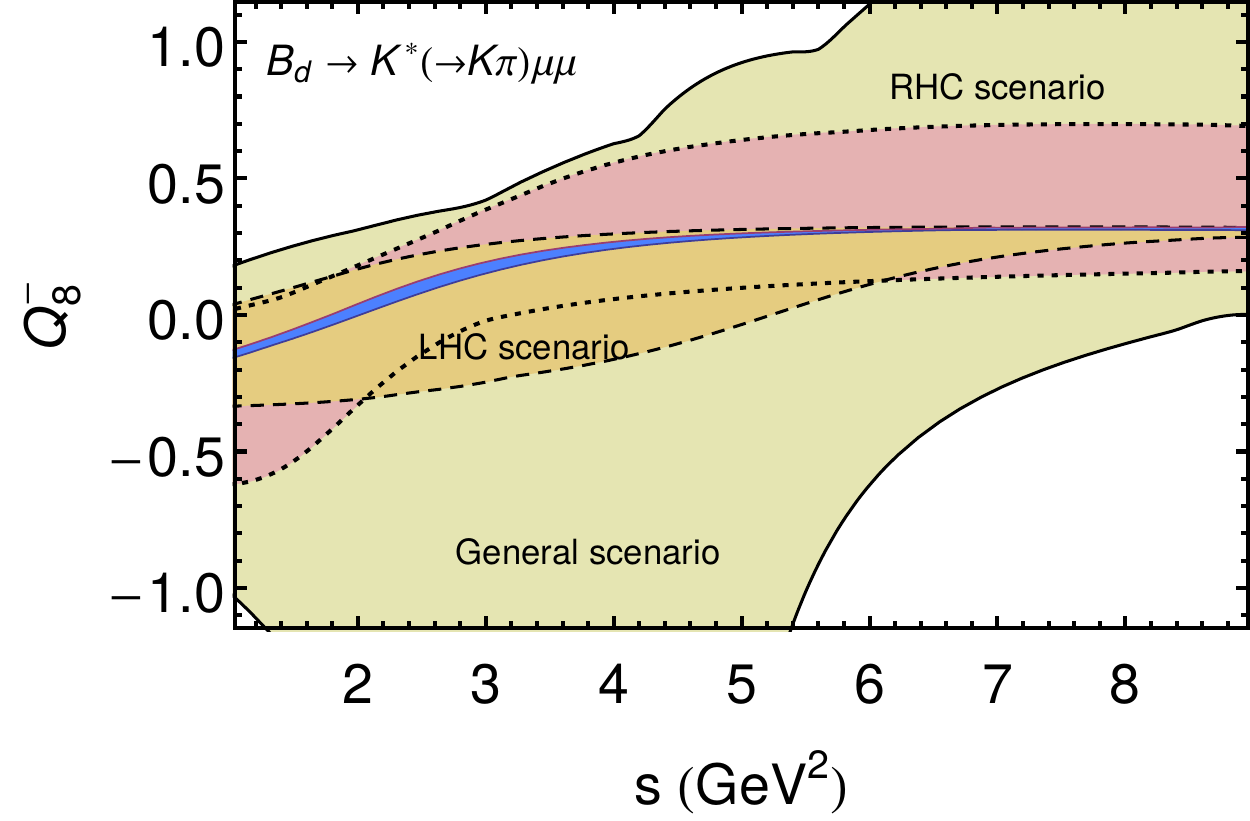}\hspace{4mm}
\includegraphics[width=7.5cm,height=5cm]{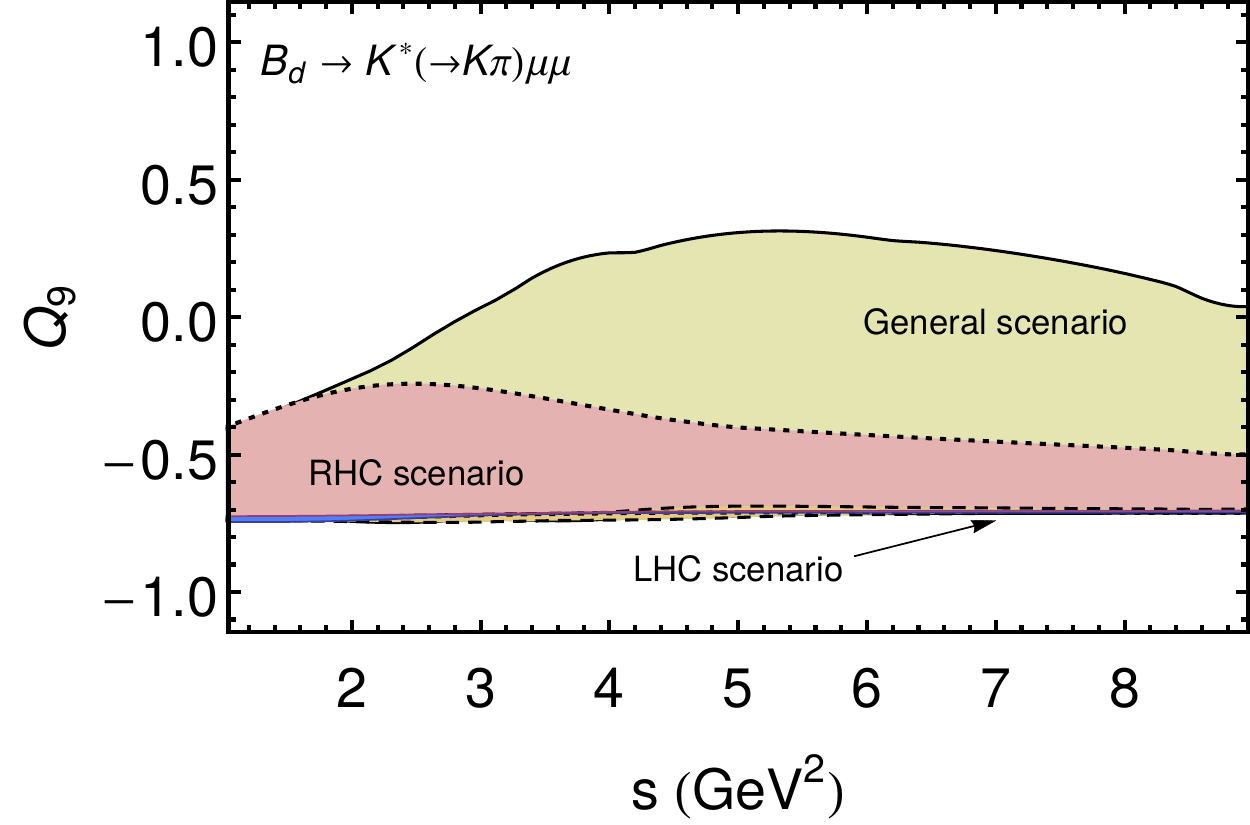}
\caption{NP reach of the observables $Q_8^-$ and $Q_9$ in the large-recoil region. See the text for details.}
\label{figNPscan}
\end{figure}

\begin{itemize}

\item LHC (Left-Handed Currents) scenario: NP contributions to $\C7,\C9,\C{10}$ only. This corresponds to the orange regions
in Fig.~\ref{figNPscan}, delimited by dashed lines (along the line $Q_9=-1$ on the right-hand plot).

\item RHC (Right-Handed Currents) scenario: NP contributions to $\C{7'},\C{9'},\C{10'}$ only.
This corresponds to the red regions in Fig.~\ref{figNPscan}, delimited by dotted lines.

\item General NP scenario: NP contributions to all six coefficients $\C{7^{(\prime)}},\C{9^{(\prime)}},\C{10^{(\prime)}}$.
This corresponds to the regions in green in Fig.~\ref{figNPscan}, with solid borders.

\end{itemize}

We also show the SM predictions for comparison (blue bands in Fig.~\ref{figNPscan}, with $Q_9^{\sss\rm SM}\simeq-1$
and $Q_9^{\sss\rm SM}\simeq-0.7$ for the $B_s$ and $B_d$ cases respectively).
We see that NP can indeed have a large impact on $Q_8^-$, $Q_9$. As discussed in Section~\ref{sec:Q8Q9},
any significant deviation of $Q_9$ from $Q_9^\text{SM}\simeq-\cos\tilde\phi_q$ requires right-handed currents.

We finish our exploratory NP analysis by studying a few motivated benchmark NP scenarios:

\begin{figure}
\includegraphics[width=7.5cm,height=5cm]{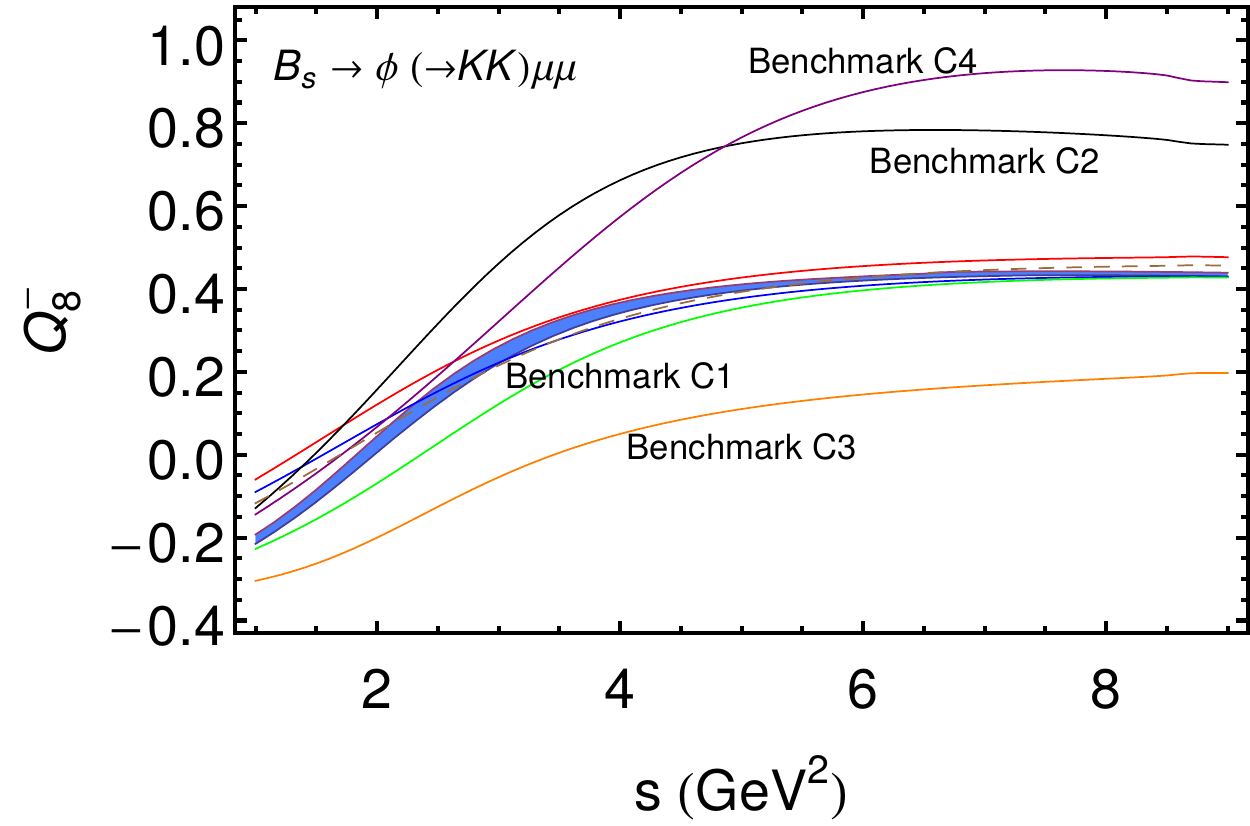}\hspace{4mm}
\includegraphics[width=7.5cm,height=5.1cm]{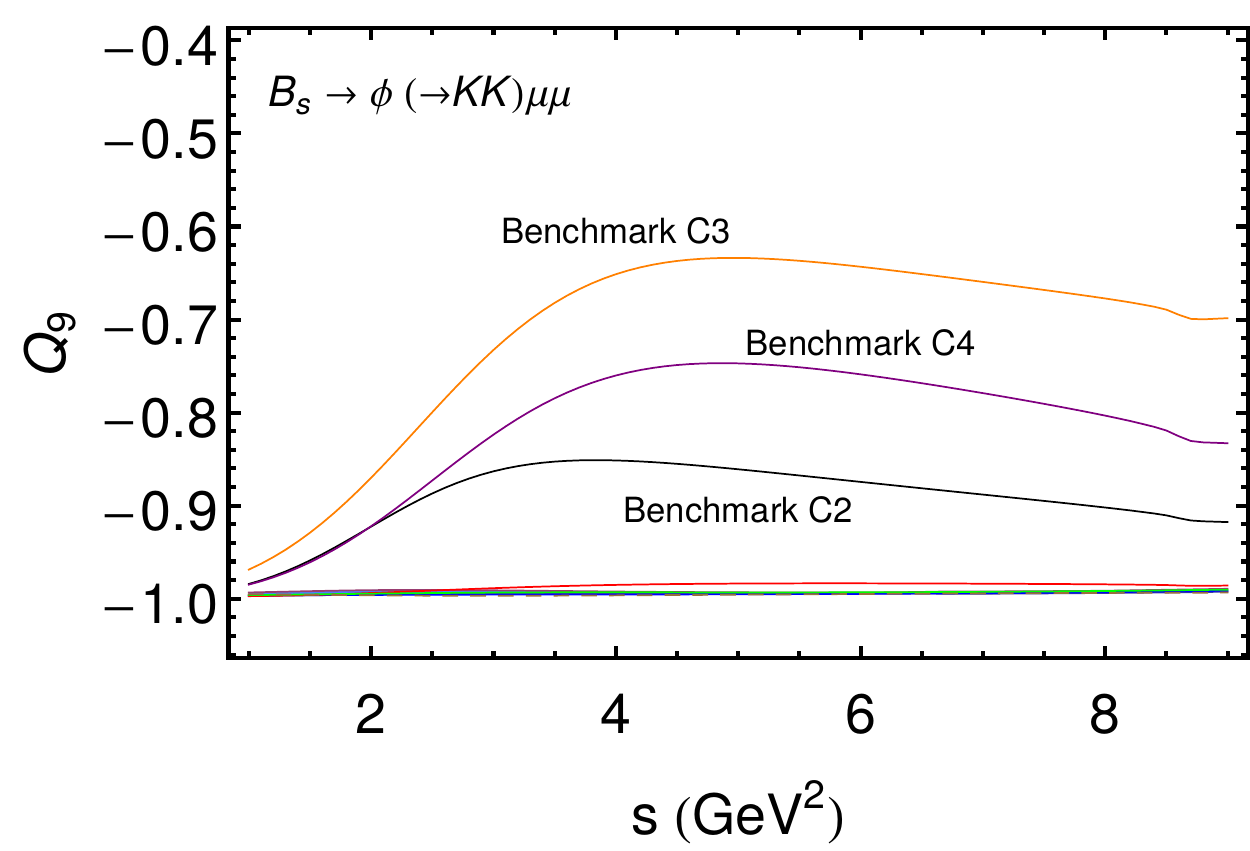}\\[4mm]
\includegraphics[width=7.5cm,height=5cm]{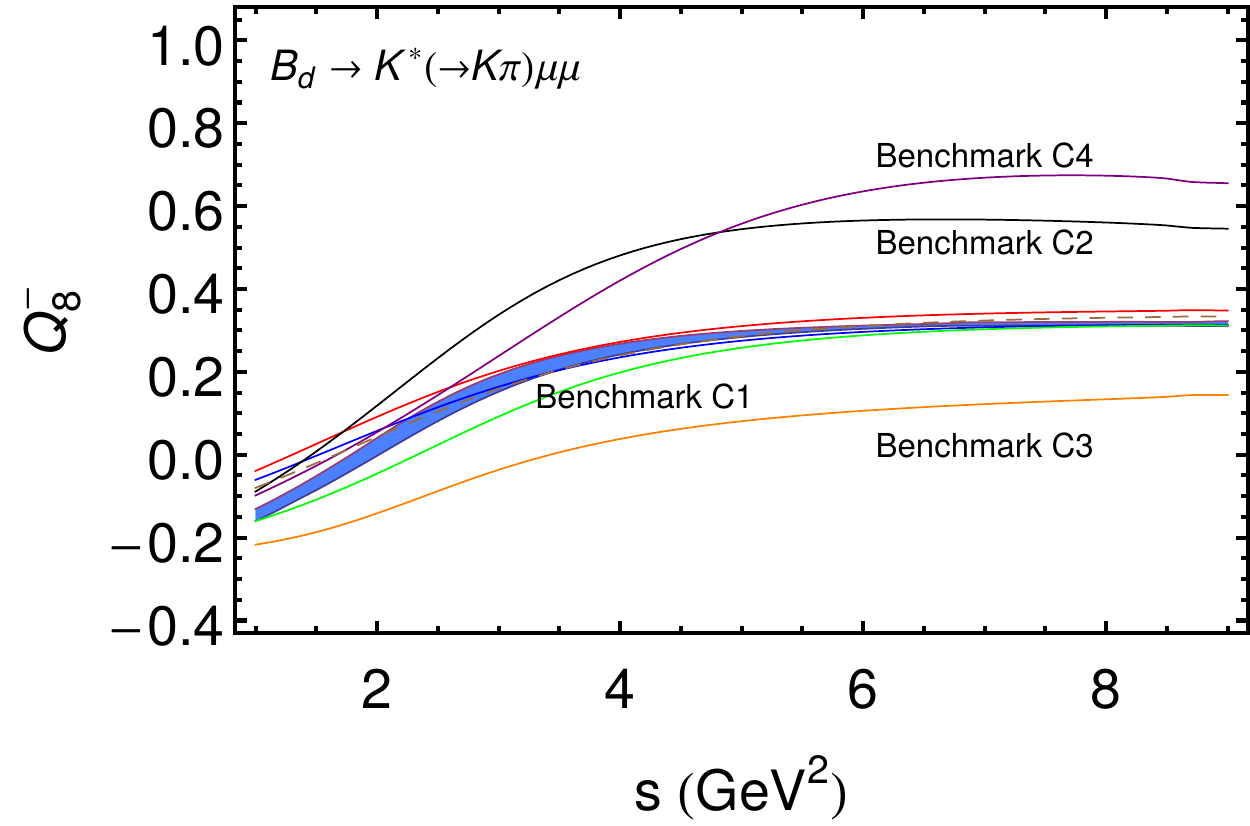}\hspace{4mm}
\includegraphics[width=7.5cm,height=5.1cm]{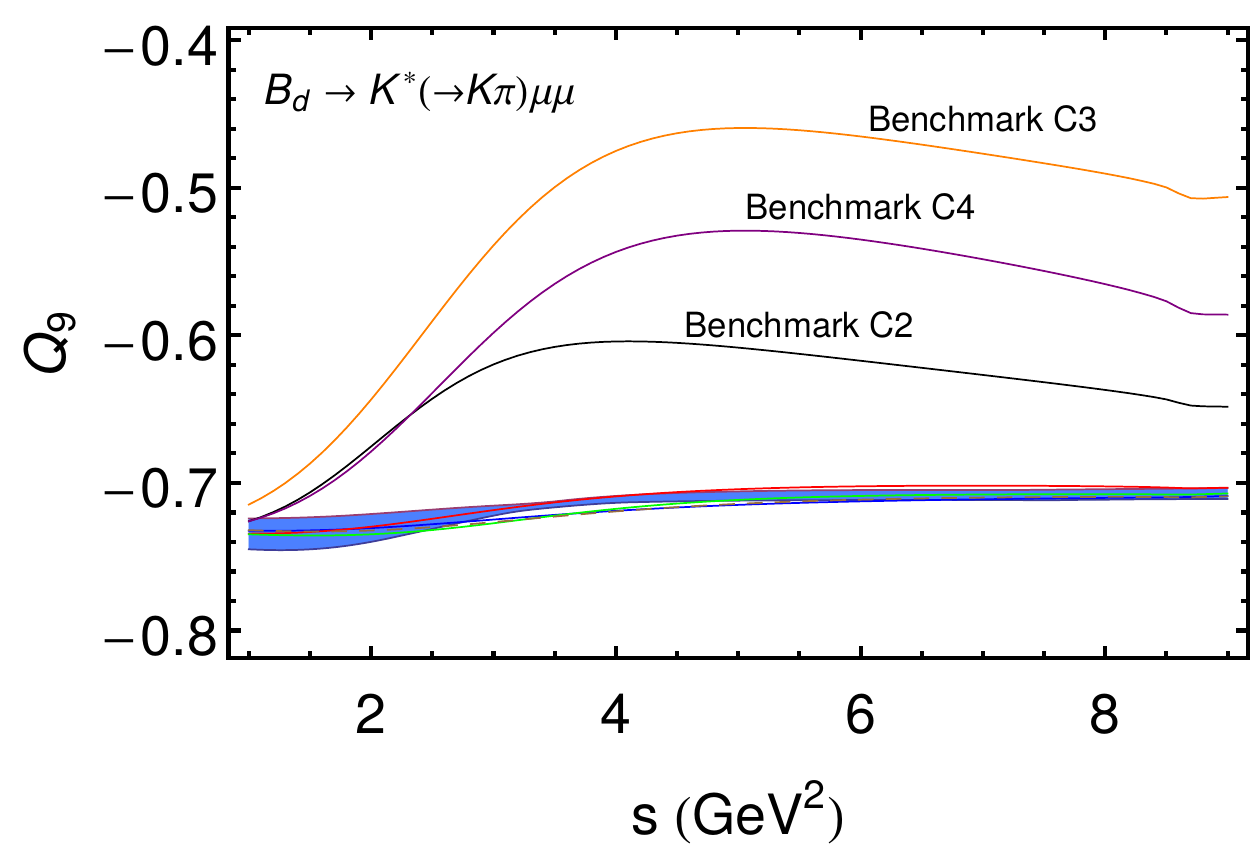}
\caption{NP benchmarks for the observables $Q_8^-$ and $Q_9$ in the large-recoil region. See the text for details.
Benchmarks A (blue), B (red) and D (dashed) are hardly visible.}
\label{figNPbench}
\end{figure}

\begin{itemize}

\item[A.] Best fit point in the $\C7-\C9$ scenario of Ref.~\cite{1307.5683}:
\eq{
\C7^{\sss\rm NP}=-0.02,\quad \C9^{\sss\rm NP}=-1.6\ .\nn
}

\item[B.] Best fit point in the $\C9-\C{9'}$ scenario of Ref.~\cite{1411.3161}:
\eq{
\C9^{\sss\rm NP}=-1.28,\quad \C{9'}^{\sss\rm NP}=0.47\ .\nn
}

\item[C.] $Z'$-motivated $\C{9^{(\prime)}},\C{10^{(\prime)}}$ scenarios (see e.g. Refs.~\cite{1211.1896,1309.2466}):

\begin{itemize}

\item[C1.] $\C9^{\sss\rm NP}=-\C{10}^{\sss\rm NP}=-1$

\item[C2.] $\C{9'}^{\sss\rm NP}=-\C{10'}^{\sss\rm NP}=1$

\item[C3.] $\C9^{\sss\rm NP}=\C{9'}^{\sss\rm NP}=-\C{10}^{\sss\rm NP}=-\C{10'}^{\sss\rm NP} = -1$

\item[C4.] $\C9^{\sss\rm NP}=-\C{9'}^{\sss\rm NP}=-\C{10}^{\sss\rm NP}=\C{10'}^{\sss\rm NP} = -1$

\end{itemize}

\item[D.] Best fit point in the general fit of Ref.~\cite{1307.5683}:
\eq{
\C7^{\sss\rm NP} = -0.02\ ,\ \C9^{\sss\rm NP} = -1.3\ ,\ \C{10}^{\sss\rm NP} = 0.3\ ,\ 
\C{7'}^{\sss\rm NP} = -0.01\ ,\ \C{9'}^{\sss\rm NP} = 0.3\ ,\ \C{10'}^{\sss\rm NP} = 0\ .\nn
}

\end{itemize}

Scenarios C.1 and C.2 arise also respectively in singlet/triplet and doublet leptoquark models
motivated by recent data on the ratio ${\cal B}(B\to K\mu\mu)/{\cal B}(B\to Kee)$ (see Ref.~\cite{1408.1627}).

The predictions for the observables $Q_8^-$ and $Q_9$ within the benchmark scenarios are shown in Fig.~\ref{figNPbench},
together with the SM prediction.
We see that, among the considered scenarios, the only ones leading to significant deviations with respect to the SM
are scenarios C (corresponding to large NP contributions to $\C{9'}$ and $\C{10'}$), while scenarios A, B and D are very close to the blue band (corresponding to the SM prediction).
As discussed before, scenario C1 has no impact on $Q_9$ as it has no right-handed currents. Therefore,
measurements of these observables compatible with the SM would give support to the best fit points obtained in
the global fits of Ref.~\cite{1307.5683,1411.3161} that we have considered, with the potential to exclude
the scenarios with $\C{9^{(\prime)}}^{\sss\rm NP},\C{10^{(\prime)}}^{\sss\rm NP}$ such as the one discussed in Ref.~\cite{1408.1627}. As an alternative
viewpoint, these observables could test the latter scenarios, and provide an alternative confirmation
if more accurate measurements for time-integrated observables happened to confirm any of them.

\section{Conclusions}
\label{sec:conc}

Decays of the type $B_{d,s}\to V(\to M_1M_2)\ell^+\ell^-$ mediated by the underlying flavour-changing neutral current process
$b\to s\ell\ell$ are of great phenomenological interest for two reasons: they lead to a vast set of independent
experimental observables, and they exhibit a remarkable sensitivity to New Physics.
The decay mode $B_d\to K^{*0}(\to K^-\pi^+)\mu^+\mu^-$ has been the first one to be carefully scrutinized,
both experimentally~\cite{0904.0770,1108.0695,1204.3933,1301.1700,1304.6325,1308.1707,1308.3409,1403.8044,1408.0978,
1501.03309}
and theoretically~\cite{1205.1438,1205.1838,1205.1845,1205.3683,1206.0273,1206.2970,1207.4004,1209.0262,1209.1525,
1210.5279,1211.6453,1212.2321,1301.7535,1305.4808,1306.3775,1307.5683,1308.1501,1308.1959,1308.4379,
1310.1082,1310.1937,1310.2478,1310.3722,1310.3887,1311.3876,1311.6729,1312.1923,1312.5267,1401.2145,1401.6707,
1402.2844,1402.6855,1403.1269,1403.2944,1405.3850,1405.5182,1406.0566,1407.6700,1411.0131,
1411.0922,411.6423,B1412.1003,1412.2955,1412.3183,1501.00367,1501.00993,1501.05193}, and first angular analyses of
the decays $B_s\to \phi(\to K^+K^-)\mu^+\mu^-$ \cite{1305.2168} and $B_d\to K^{*0}(\to K^-\pi^+)e^+e^-$~\cite{1501.03038}
have been already performed.

In the case where $M_1M_2$ is a CP eigenstate (such as $B_d\to K^{*0}(\to K_S\pi^0)\ell^+\ell^-$,
$B_s\to \phi(\to K^+K^-)\mu^+\mu^-$ or $B_s\to \phi(\to K_SK_L)\mu^+\mu^-$), neutral $B$-meson mixing interferes with the decay,
leading to interesting differences with respect to flavour-specific processes where mixing plays no role (such as
$B_d\to K^{*0}(\to K^-\pi^+)\mu^+\mu^-$). In this paper we have studied the effects induced by neutral-meson mixing
for the analysis of exclusive $B\to V\ell\ell$ decays, spelling
out the theoretical formalism and analysing its phenomenological consequences.

As a first observation, the angular distributions become time-dependent, with additional structures
compared to the case without mixing. These structures are the new angular coefficients $h_i$ and $s_i$,
defined in Eqs.~(\ref{eq:J+Jt}),(\ref{eq:J-Jt}) and given explicitly in terms of the different amplitudes
in Appendix~\ref{app2}. Two types of observables can then be defined for these modes:
time-integrated and time-dependent observables. The first type depend on the experimental set-up
($B$-factory or hadronic machine) and differ from the corresponding observables in decays
without mixing by multiplicative factors depending on the mixing parameters $x$ and $y$. In addition, the expressions
for time-integrated observables at hadronic machines include an extra term proportional to the
coefficients $h_i$ or $s_i$. This is similar to analogous relations derived for
$B_s\to \phi(\to K^+K^-)\mu^+\mu^-$~\cite{0805.2525}, $B_s\to VV$~\cite{1111.4882} and
$B_s\to \mu^+\mu^-$~\cite{1204.1737}. The corresponding expressions for time-integrated
observables are given in Eqs.~(\ref{eq:<J+Jt>Had})-(\ref{eq:<J-Jt>Bfac}). However, it seems difficult to
extract $h_i$ or $s_i$ using time-integrated observables, as they are suppressed by small meson-mixing parameters.

On the other hand, a time-dependent angular analysis with flavour tagging paves the way for the
observables $s_i$. We identify $s_8$ and $s_9$ as the most interesting observables, as they are expected to be large even
in the absence of CP violation.
We have demonstrated that these observables contain new information compared to the angular coefficients $J_i$,
and we have built ``optimised" versions of these observables with reduced sensitivity to form factors.
We have focused on two such observables, called $Q_8^-$ and $Q_9$ and defined in Eqs.~(\ref{Q8m}),(\ref{Q9}).
These observables can be predicted in the SM with good precision (see Fig.~\ref{figQ8Q9}), and show good sensitivity
to particular New Physics scenarios (see Figs.~\ref{figNPscan} and~\ref{figNPbench}).

Current analyses of $b\to s\ell\ell$ transitions point towards deviations compared to SM expectations, explained via
large NP contributions to $\C9$ (and potentially smaller contributions to other Wilson coefficients). It is particularly
interesting and useful to cross check this trend from other sources. Our analysis shows that additional information
could come from time-dependent angular analyses of tagged $B_{d,s}\to V(\to M_1M_2)\ell\ell$ decays,
with $M_1M_2$ a CP eigenstate. We thus encourage exploratory studies to determine the experimental feasibility of
such analyses, in particular in the context of a high-luminosity flavour factory such as Belle-II.

\section*{Acknowledgements}
We thank Damir Becirevic, Emi Kou, Quim Matias and Dirk Seidel for useful discussions.
JV acknowledges the hospitality at LPT-Orsay where part of this work was done.
JV is funded by the DFG within research unit FOR 1873 (QFET).


\newpage

\appendix

\section{CP-conjugate kinematics from invariants}
\label{app1}

The kinematics of the four-body decay $B\to V[\to M_1(p_1) M_2(p_2)] \ell^+(p_+)\ell^-(p_-)$, is completely specified by
four invariant masses (e.g. $s_{+-}\equiv 2p_+\cdot p_-$, $s_{1+}\equiv 2p_1\cdot p_+$, $s_{1-}\equiv 2p_1\cdot p_-$
and $s_{2+}\equiv 2p_2\cdot p_+$), and the sign of
$\epsilon_{12+-}\equiv \epsilon_{\mu\nu\rho\lambda} p_1^\mu p_2^\nu p_+^\rho p_-^\lambda$,
which defines the parity of the final state \cite{1411.7677}.
We set these kinematic invariants to some fixed values $s,s_1,s_2,s_3,\sigma$ such that:
\eq{
s_{+-} = s,\ \ s_{1+}=s_1,\ \ s_{1-}=s_2,\ \ s_{2+}=s_3,
\ \ \text{sgn}(\epsilon_{12+-}) =\sigma\ .
}

The kinematics of the CP-conjugated decay
$\bar B\to \bar V[\to \bar M_1(p_{\bar 1}) \bar M_2(p_{\bar 2})] \ell^+(p_+)\ell^-(p_-)$
is specified analogously. Under the condition of CP-conservation, the differential rates of both
decays ($d\Gamma$ and $d\bar \Gamma$) must be equal at CP-conjugated kinematic points:
\eqa{
d\Gamma \big(s_{+-} = s,s_{1+}=s_1,s_{1-}=s_2,s_{2+}=s_3,\text{sgn}(\epsilon_{12+-}) =\sigma\big) &&
\label{CPcorr}\\[2mm]
&&\hspace{-10.5cm} = d\bar\Gamma \big(s_{+-} = s,s_{\bar1-}=s_1,s_{\bar1+}=s_2,s_{\bar2-}=s_3,
\text{sgn}(\epsilon_{\bar1\bar2-+}) =-\sigma\big)\nn\\[2mm]
&&\hspace{-10.5cm} = d\bar\Gamma \big(s_{+-} = s,s_{\bar1+}=s_2,s_{\bar1-}=s_1,s_{\bar 2+}=m_B^2-m_V^2-s-s_1-s_2-s_3,
\text{sgn}(\epsilon_{\bar1\bar2+-}) =\sigma\big)\ ,\nn
}
where we have made the replacements $\{1,2,+,-,\}\to \{\bar1,\bar2,-,+\}$ and $\sigma\to-\sigma$ to account for
$C$ and $P$ transformations respectively. In addition, we have used momentum conservation and neglected light-meson and
lepton masses to write
\eq{
s_{\bar2 -} = m_B^2 - m_V^2 - s_{+-} - s_{\bar1 +} - s_{\bar1 -} - s_{\bar2 +}\ .
}

The angular distribution is obtained by expressing these rates in terms of $s_{+-}$, two polar angles
$\theta_M, \theta_\ell$ ($0<\theta_i<\pi$), and one azimuthal angle\footnote{
The kinematic angle $\phi$ should not be confused with the mixing angle, not appearing in this appendix.}
$\phi$  ($0<\phi<2\pi$).
In the case of $d\Gamma$, the angles are usually defined as \cite{9907386}

\begin{itemize}

\item $\theta_M=\theta_1$: Polar angle between the momenta $\vec p_B$ and $\vec p_1$ in the rest-frame of $V$.
In terms of momentum invariants, we find:
\eq{
\cos\theta_1 = \frac{m_B^2-m_V^2-s_{+-} - 2 s_{1+} - 2 s_{1-}}{\sqrt{(m_B^2-m_V^2 -s_{+-})^2 - 4m_V^2 s_{+-}}}
\equiv c_\theta(s_{+-},s_{1+},s_{1-})\ ,
\label{theta1}}
and $\sin\theta_1 = + \sqrt{1-\cos\theta_1^2}$ by definition.

\item $\theta_\ell=\theta_+$: Polar angle between the momenta $\vec p_B$ and $\vec p_+$ in the dilepton rest-frame.
In terms of momentum invariants, we find:
\eq{
\cos\theta_+ = \frac{m_B^2-m_V^2-s_{+-} - 2 s_{1+} - 2 s_{2+}}{\sqrt{(m_B^2-m_V^2 -s_{+-})^2 - 4m_B^2 s_{+-}}}
\equiv c_\theta(s_{+-},s_{1+},s_{2+})\ ,
\label{theta+}}
and again $\sin\theta_+ = + \sqrt{1-\cos\theta_+^2}$.

\item $\phi=\phi_{12+-}$: Oriented angle between the planes specified by $(\vec p_1,\vec p_2)$ and $(\vec p_+,\vec p_-)$ in the $B$-meson
rest frame. The orientation is specified by (in this frame):
\eq{
\cos\phi = \frac{(\vec p_1 \times \vec p_2)\cdot (\vec p_+ \times \vec p_-)}
{|\vec p_1 \times \vec p_2|\cdot|\vec p_+ \times \vec p_-|}\ ,\quad
\sin\phi = \frac{(\vec p_++\vec p_-)\cdot [(\vec p_1 \times \vec p_2)\times (\vec p_+ \times \vec p_-)]}
{|\vec p_++\vec p_-|\cdot|\vec p_1 \times \vec p_2|\cdot|\vec p_+ \times \vec p_-|}\ ,
}
which in terms of momentum invariants gives:
\eq{
\cos\phi_{12+-} = c_\phi(s,s_{1+},s_{1-},s_{2+})\ ,\quad
\sin\phi_{12+-} = \text{sgn}(\epsilon_{12+-}) \sqrt{1- c_\phi^2}\ ,
}
where $c_\phi(s,s_{1+},s_{1-},s_{2+})= a/(2 m_V\,b\,c)$ with
\eqa{
a &=&  m_V^4 (s + s_{1+}) - (m_B^2 - s) \big[s_{1+} (-m_B^2 + s + s_{1-} + s_{1+}) + (s_{1-} + s_{1+}) s_{2+}\big] + \nn\\[2mm]
&& m_V^2\,\big[s^2 - 2 s_{1+} + (s_{1-} + s_{1+}) (s_{1+} + s_{2+}) - s\,(m_B^2 - 2 s_{1-} - 2 s_{1+} - 2 s_{2+})\big]\ , \nn\\[2mm]
b &=& \sqrt{s_{1-} + s_{1+} - (m_V^2 + s_{1-} + s_{1+}) (s + s_{1-} + s_{1+})}\ ,\nn\\[2mm]
c&=& \sqrt{s\,\big[(s_{1+} + s_{2+}) (m_B^2 - s - s_{1+} - s_{2+}) - m_V^2 (s + s_{1+} + s_{2+})\big]}\ .
}

We note that $\sin\phi$ is proportional to $\text{sign}(\epsilon_{12+-})$.

\end{itemize}

Other possibilities are $\theta_M=\theta_2$ or $\theta_\ell=\theta_-$, obtained from
(\ref{theta1}), (\ref{theta+}) by obvious replacements. In the case of the CP-conjugate mode,
the angles $\theta_{\bar1,\bar2}$, $\theta_{\pm}$, $\phi_{\bar i\bar j\pm\mp}$
are defined analogously.

In terms of the angular distribution, the CP correspondence in Eq.~(\ref{CPcorr}) depends on how the angles are defined
for $d\bar \Gamma$, relative to $d\Gamma$. We recall that in the case of flavour-specific (``self-tagging") modes,
such as $B_d\to K^*(\to K^+\pi^-)\ell\ell$, one might choose any convention for $d\Gamma$ and $d\bar\Gamma$ independently,
as the final states are different and distinguishable. However, this is not the case for untagged flavour-non-specific decays,
where the final states arising from the $B$ and the $\bar B$ decay cannot be distinguished.
We consider three different conventions:

\begin{itemize}

\item[\bf A.] \underline{$d\Gamma(s,\theta_1,\theta_+,\phi_{12+-})$ and $d\bar\Gamma(s,\theta_{\bar1},\theta_+,\phi_{\bar1\bar2+-})$:}
This is the usual theory convention in $B_d\to K^*\ell\ell$, where in \emph{both}
CP-conjugated modes $\theta_M$ is defined with respect to the kaon, $\theta_\ell$ with respect
to the positively-charged lepton, and the orientation of $\phi$ is given by $\phi_{K\pi +-}$.
This is also the only possible convention in untagged decays with $\bar M_1 = M_1$ and $\bar M_2=M_2$, such as
$B_d\to K^*(\to K_S \pi^0)\ell^+\ell^-$ and $B_s\to \phi(\to K_SK_L)\ell^+\ell^-$ at hadronic machines.
With this convention, Eq.~(\ref{CPcorr}) implies
\eqa{
&&d\Gamma \big(s_{+-} = s,\cos\theta_1 = c_\theta(s,s_1,s_2),\cos\theta_+ = c_\theta(s,s_1,s_3),\\
&&\hspace{3cm}\cos\phi = c_\phi(s,s_1,s_2,s_3),\text{sgn}(\sin\phi)=\sigma\big) = \nn\\[2mm]
&&d\bar\Gamma \big(s_{+-} = s,\cos\theta_{\bar1} = c_\theta(s,s_2,s_1),\cos\theta_+ = c_\theta(s,s_2,m_B^2-m_V^2-s-s_1-s_2-s_3),\nn\\
&&\hspace{3cm}\cos\phi = c_\phi(s,s_2,s_1,m_B^2-m_V^2-s-s_1-s_2-s_3),\text{sgn}(\sin\phi)=\sigma\big)\nn
}
We note the following relations:
\eqa{
c_\theta(s,s_1,s_2) \equiv X &=& c_\theta(s,s_2,s_1) \ ,\\
c_\theta(s,s_1,s_3) \equiv Y &=& -c_\theta(s,s_2,m_B^2-m_V^2-s-s_1-s_2-s_3)\ ,\\
c_\phi(s,s_1,s_2,s_3) \equiv Z &=& -c_\phi(s,s_2,s_1,m_B^2-m_V^2-s-s_1-s_2-s_3)\ .
}
Therefore,
\eqa{
&&d\Gamma(s_{+-}=s,\cos\theta_1=X, \cos\theta_+=Y, \cos\phi = Z, \text{sgn}(\sin\phi)=\sigma) = \nn\\[2mm]
&&\hspace{1cm}d\bar\Gamma (s_{+-}=s,\cos\theta_{\bar1}=X, \cos\theta_+=-Y, \cos\phi = -Z, \text{sgn}(\sin\phi)=\sigma)\ .
\quad
}
With the angles defined in this way, the two angular distributions are written as:
\eq{
d\Gamma = \sum_i J_i(s) f_i(\theta_\ell,\theta_M,\phi)\ ,\quad d\bar\Gamma =
\sum_i \zeta_i \bar J_i(s) f_i(\theta_\ell,\theta_M,\phi)\ ,
\label{case1}}
with $\zeta_{1,2,3,4,7}=1\ , \ \zeta_{5,6,8,9}=-1$.

\item[\bf B.] \underline{$d\Gamma(s,\theta_1,\theta_+,\phi_{12+-})$ and $d\bar\Gamma(s,\theta_{\bar1},\theta_-,\phi_{\bar1\bar2+-})$:}
This is the usual experimental convention for $B_d\to K^*\ell\ell$, where in
both modes $\theta_M=\theta_K$ and $\phi = \phi_{K\pi +-}$, but for $\theta_\ell$ one takes $\ell^+$ or $\ell^-$
for the $B$ and $\bar B$ decay respectively. We have:
\eqa{
&&d\Gamma \big(s_{+-} = s,\cos\theta_1 = c_\theta(s,s_1,s_2),\cos\theta_+ = c_\theta(s,s_1,s_3),\\
&&\hspace{3cm}\cos\phi = c_\phi(s,s_1,s_2,s_3),\text{sgn}(\sin\phi)=\sigma\big) = \nn\\[2mm]
&&d\bar\Gamma \big(s_{+-} = s,\cos\theta_{\bar1} = c_\theta(s,s_2,s_1),
\cos\theta_- = c_\theta(s,s_1,s_3),\nn\\
&&\hspace{3cm}\cos\phi = c_\phi(s,s_1,s_2,s_3),\text{sgn}(\sin\phi)=\sigma\big)\nn
}
which means that, with this convention,
\eq{
d\Gamma = \sum_i J_i(s) f_i(\theta_\ell,\theta_M,\phi)\ ,\quad d\bar\Gamma =
\sum_i \bar J_i(s) f_i(\theta_\ell,\theta_M,\phi)\ .
}

\item[\bf C.] \underline{$d\Gamma(s,\theta_1,\theta_+,\phi_{12+-})$ and $d\bar\Gamma(s,\theta_{\bar2},\theta_+,\phi_{\bar2\bar1+-})$:}
This is the only possible convention for the case of untagged decays where $\bar M_1 = M_2$ and $\bar M_2=M_1$,
as for example the decay $B_s\to \phi(\to K^+K^-)\ell^+\ell^-$ at a hadronic machine. In this case,
\eqa{
&&\hspace{-5mm}d\Gamma \big(s_{+-} = s,\cos\theta_1 = c_\theta(s,s_1,s_2),\cos\theta_+ = c_\theta(s,s_1,s_3),\nn\\
&&\hspace{2.5cm}\cos\phi = c_\phi(s,s_1,s_2,s_3),\text{sgn}(\sin\phi)=\sigma\big) = \nn\\[2mm]
&&\hspace{-5mm}d\bar\Gamma \big(s_{+-} = s,\cos\theta_{\bar2} = c_\theta(s,m_B^2-m_V^2-s-s_1-s_2-s_3,s_3),\\
&&\hspace{2.5cm}\cos\theta_+ = c_\theta(s,s_2,m_B^2-m_V^2-s-s_1-s_2-s_3),\nn\\
&&\hspace{2.5cm}\cos\phi = c_\phi(s,m_B^2-m_V^2-s-s_1-s_2-s_3,s_3,s_2),\text{sgn}(\sin\phi)=-\sigma\big)\nn\ .
}
Using the relations
\eqa{
c_\theta(s,s_1,s_2) \equiv X &=& -c_\theta(s,m_B^2-m_V^2-s-s_1-s_2-s_3,s_3) \ ,\\
c_\theta(s,s_1,s_3) \equiv Y &=& -c_\theta(s,s_2,m_B^2-m_V^2-s-s_1-s_2-s_3)\ ,\\
c_\phi(s,s_1,s_2,s_3) \equiv Z &=& +c_\phi(s,m_B^2-m_V^2-s-s_1-s_2-s_3,s_3,s_2)\ ,
}
we have
\eqa{
&&d\Gamma(s_{+-}=s,\cos\theta_1=X, \cos\theta_+=Y, \cos\phi = Z, \text{sgn}(\sin\phi)=\sigma) = \nn\\[2mm]
&&\hspace{1cm}d\bar\Gamma (s_{+-}=s,\cos\theta_{\bar2}=-X, \cos\theta_+=-Y, \cos\phi = Z, \text{sgn}(\sin\phi)=-\sigma)\ .
\quad
}
With this convention, the differential rates are given by:
\eq{
d\Gamma = \sum_i J_i(s) f_i(\theta_\ell,\theta_M,\phi)\ ,\quad d\bar\Gamma =
\sum_i \zeta_i \bar J_i(s) f_i(\theta_\ell,\theta_M,\phi)\ ,
\label{case3}}
with $\zeta_{1,2,3,4,7}=1,\ \zeta_{5,6,8,9}=-1$. This is the same as Eq.~(\ref{case1}) but for a different reason.

\end{itemize}

We see that conventions A and C yield the same relation between $d\Gamma$ and $d\bar\Gamma$, but they apply to
different kinds of modes. For decays into flavour-specific modes, convention B is also possible, but with a different
relationship between $d\Gamma$ and $d\bar\Gamma$. In the present paper, we choose convention A for decays
into flavour-specific modes as well as for $B\to V(\to M_1M_2)\ell\ell$ decays with $\bar M_1=M_1$, $\bar M_2=M_2$,
and convention C for $B\to V(\to M_1M_2)\ell\ell$ decays with $\bar M_1=M_2$, $\bar M_2=M_1$.

\section{CP-parities associated to transversity amplitudes}
\label{appCP}

We consider the decay $B\to V N$ (cf.~Eq~(\ref{eq:BVN})) where $V$ and $N$ are unstable particles,
$V$ decaying into two particles $M_1$ and $M_2$. As shown in Ref.~\cite{Dunietz:1990cj}, the CP-parity of a final state
$X$ is given by
\begin{equation}
\eta_X=\xi (-1)^\tau
\end{equation}
with $\tau=\tau(M_1)+\tau(M_2)+\tau(N)$ the ``transversity" of the state $M_1M_2N$ (defined below),
and $\xi$ depends on the class of decay:
\begin{itemize}
\item class 1: $V$ (not necessarily with a definite spin) decays into $M_1$ and $M_2$ which are CP-eigenstates,
and $N$ decays into a CP-eigenstate. In this case,
\eq{
\xi=\eta(N)\eta(M_1)\eta(M_2).
}

\item class 2: $V$ (with a definite spin $s_V$) decays into spin-0 $M_1$ and $M_2$ which are CP-conjugates,
and $N$ decays into a CP-eigenstate. In this case,
\eq{
\xi=\eta(N)(-1)^{s_V}.
}

\item class 3: $V$ and $N$ are CP-conjugates with a definite spin $s_V$, with $V$ decaying into spin-0 $M_1$ and $M_2$.
In this case,
\eq{
\xi=(-1)^{s_V}.
}

\end{itemize}
Here $\eta(V)$ and $s_V$ are the intrinsic CP-parity and spin of the particle $V$.
The first class is illustrated by the time-dependent analysis of
$B_d\to J/\psi K^*(\to K_S\pi^0)$~\cite{Aubert:2004cp}.
For the class-1 processes ($B_d\to K^*(\to K_S\pi^0)\ell\ell$
and $B_s\to \phi(\to K_SK_L)\ell\ell$) and class-2 process ($B_s\to \phi(\to K^+K^-)\ell\ell$) of interest, we have
\begin{equation}
\eta_X=\eta(N)(-1)^{\tau(N)+1}\eta
\end{equation}
where $\eta=-\eta(M_1)\eta(M_2)$ for class-1, and $\eta=1$ for class-2.
For all the processes considered here, the combinations of intrinsic CP-parities yield $\eta=1$.

\begin{table}
\begin{center}
\begin{tabular}{cccccc}
$X$              & $N$ & $\eta(N)$ & $s$ & $\tau(N)$ & $\eta_X$\\
\hline
${0}$ &  vector $\gamma^*,Z$ (or axial) with $\epsilon_{0}$ & 1 & 1 & 1 & $\eta$\\
${||}$ &  vector $\gamma^*,Z$ (or axial) with $\epsilon_{||}$ & 1 & 1 & 1 & $\eta$\\
${\perp}$ & vector $\gamma,Z$ (or axial) with $\epsilon_{\perp}$ & 1 & 1& 0 & $-\eta$\\
$S $ & scalar $H$ & 1 & 1 & 0 & $-\eta$\\
$t$ & vector $\gamma^*,Z$ (or axial) with $\epsilon_{t}$ & -1 & 0 & 0 & $\eta$\\
$t$ & pseudoscalar $A$ & -1 & 0 & 0 & $\eta$\\
\hline
\end{tabular}
\end{center}
\caption{Properties of the transversity amplitudes involved in $B\to V\ell\ell$:
CP-parity of $N$, spin of the lepton pair, transversity of $N$, and CP-parity of the final state.
 }\label{tab:CPparity}
\end{table}

In order to determine the CP-parity of the different transversity states, we have thus to determine $\eta(N)$ and $\tau(N)$:

\begin{itemize}

\item Using the language of Ref.~\cite{Dunietz:1990cj}, we see that $A_{0},A_{\perp},A_{||},A_{S,t}$ are respectively associated with the combinations 
of helicity amplitudes denoted ${\cal G}^{1+}_{0,0,0}, {\cal G}^{1+}_{1,0,0}, {\cal G}^{1-}_{1,0,0}, {\cal G}^{0+}_{0,0,0}$, with respective CP parities 
$-\xi,-\xi,\xi,\xi$. For our decays, it implies that we should have the following associations (modulo 2):
\begin{equation}
A_0,A_{||}: \tau=1, \qquad A_\perp,A_t,A_S: \tau=0
\end{equation}
The states corresponding to different transversities can be accessed through the angular analysis of the decay.

\item $\eta(N)$ can be determined from the assumed quantum numbers of the intermediate boson (spin, polarisation, parity). $\eta(N)$ is identical for spin-1 particles with vector or axial couplings, whereas scalar and pseudoscalar $N$ have opposite CP-parities. One can also notice that 
this constrains the spin of the emitted $\ell^+\ell^-$ pair. If we denote its angular momentum $l$ and total spin $s$ (either 0 or 1), the $P$-parity of such a fermion-antifermion pair is  given by $(-1)^{l+1}$, its $C$ parity by $(-1)^{l+s}$, so that its CP-parity is $(-1)^{s+1}$. Since we assumed that the decay $N\to \ell^+\ell^-$ conserves CP-parity, we have $\eta(N)=(-1)^{s+1}$. We have $l=s=0$ corresponding to a pseudoscalar $N$, $l=s=1$ corresponding to a scalar $N$, $l=0,s=1$ corresponding to a (real) vector/axial $N$, $l=1,s=0$ corresponding to a time-like vector/axial $N$ (this can be checked from the CP-parity of the corresponding fermion-antifermion currents).

\end{itemize}

For each amplitude, we can determine the intermediate virtual boson $N$ with the appropriate quantum numbers, the corresponding spin of the lepton pair,
the transversity associated, and the CP-parity of the final state, as indicated in Table~\ref{tab:CPparity}.
We see in particular that we agree with the assignments for the class-1 decay $B_d\to J/\psi K^*(\to K_S\pi^0)$~\cite{Aubert:2004cp}.
In the end, we have
\begin{equation}
\eta_X=\eta \quad \text{for}\quad  X=L0,L||,R0,R||,t \quad; \quad \eta_X=-\eta \quad \text{for}\quad X=L\perp,R\perp,S\ .
\end{equation}
We impose $\tau(N)=0$ for a vector $N$ with timelike polarisation, to obtain the same
CP-parity as in the pseudoscalar case. This agrees with the expectation that $A_t$ should have the same CP-parity as $A_0$.

\section{Expressions for the coefficients $s_i$ and $h_i$}
\label{app2}

The coefficients $s_i$ are given by

\allowdisplaybreaks{
\begin{eqnarray}
s_{1s}&=&\frac{2+\beta_\ell^2}{2}{\rm Im}[e^{i\phi}\{\widetilde{A}^{L}_{\perp}A^{L*}_{\perp}+\widetilde{A}^{L}_{||}A^{L*}_{||}+\widetilde{A}^{R}_{\perp}A^{R*}_{\perp}+\widetilde{A}^{R}_{||}A^{R*}_{||}\}]\\
\nonumber&&\qquad
  +\frac{4m_\ell^2}{q^2}{\rm Im}[e^{i\phi}\{\widetilde{A}^{L}_{\perp}A^{R*}_{\perp}+\widetilde{A}^{L}_{||}A^{R*}_{||}\}-e^{-i\phi}\{A^{L}_{\perp}\widetilde{A}^{R*}_{\perp}+A^{L}_{||}\widetilde{A}^{R*}_{||}\}]\\
s_{1c}&=&2{\rm Im}[e^{i\phi}\{\widetilde{A}^{L}_0 A^{L*}_0+\widetilde{A}^{R}_0 A^{R*}_0\}]\\
\nonumber&&\qquad
    +\frac{8m_\ell^2}{q^2}\Big[{\rm Im}[e^{i\phi}\{\widetilde{A}_t  A^*_t\}]    +{\rm Im}[e^{i\phi}\widetilde{A}^{L}_0A^{R*}_0-e^{-i\phi}A^{L}_0\widetilde{A}^{R*}_0]\Big]+2\beta_\ell^2{\rm Im}[e^{i\phi}\widetilde{A}_SA^{*}_S]\\
s_{2s}&=&\frac{\beta_\ell^2}{2}{\rm Im}[e^{i\phi}\{\widetilde{A}^{L}_{\perp}A^{L*}_{\perp}+\widetilde{A}^{L}_{||}A^{L*}_{||}+\widetilde{A}^{R}_{\perp}A^{R*}_{\perp}+\widetilde{A}^{R}_{||}A^{R*}_{||}\}]\\
s_{2c}&=&-2\beta_\ell^2{\rm Im}[e^{i\phi}\{\widetilde{A}^{L}_0 A^{L*}_0+\widetilde{A}^{R}_0 A^{R*}_0\}]\\
s_3 &=&\beta_\ell^2{\rm Im}[e^{i\phi}\{\widetilde{A}^{L}_{\perp}A^{L*}_{\perp}-\widetilde{A}^{L}_{||}A^{L*}_{||}+\widetilde{A}^{R}_{\perp}A^{R*}_{\perp}-\widetilde{A}^{R}_{||}A^{R*}_{||}\}]\\
s_4&=&\frac{1}{\sqrt{2}}\beta_\ell^2{\rm Im}[e^{i\phi}\{\widetilde{A}^{L}_0A^{L*}_{||}+\widetilde{A}^{R}_0A^{R*}_{||}\}-e^{-i\phi}\{A^{L}_0\widetilde{A}^{L*}_{||}+A^{R}_0\widetilde{A}^{R*}_{||}\}]\\
s_5&=&\sqrt{2}\beta_\ell\Bigg[
{\rm Im}[e^{i\phi}\{\widetilde{A}^{L}_{0}A^{L*}_{\perp}-\widetilde{A}^{R}_{0}A^{R*}_{\perp}\}-e^{-i\phi}\{A^{L}_{0}\widetilde{A}^{L*}_{\perp}-A^{R}_{0}\widetilde{A}^{R*}_{\perp}\}]\\
\nonumber&&\qquad
-\frac{m_\ell}{\sqrt{q^2}}
{\rm Im}[e^{i\phi}\{\widetilde{A}^{L}_{||}A^*_S+\widetilde{A}^{R}_{||}A^*_S\}-e^{-i\phi}\{A^{L}_{||}\widetilde{A}^*_S+A^{R}_{||}\widetilde{A}^*_S\}]
\Bigg]\\
s_{6s}&=&2\beta_\ell
{\rm Im}[e^{i\phi}\{\widetilde{A}^{L}_{||}A^{L*}_{\perp}-\widetilde{A}^{R}_{||}A^{R*}_{\perp}\}-e^{-i\phi}\{A^{L}_{||}\widetilde{A}^{L*}_{\perp}-A^{R}_{||}\widetilde{A}^{R*}_{\perp}\}]\\
s_{6c}&=&4\beta_\ell\frac{m_\ell}{\sqrt{q^2}}{\rm Im}[e^{i\phi}\{\widetilde{A}^{L}_0A^*_S+\widetilde{A}^{R}_0A^*_S\}-e^{-i\phi}\{A^{L}_0\widetilde{A}^*_S+A^{R}_0\widetilde{A}^*_S\}]\\
s_7&=&-\sqrt{2}\beta_\ell \Bigg[{\rm Re}[e^{i\phi}\{\widetilde{A}^{L}_{0}A^{L*}_{||}-\widetilde{A}^{R}_{0}A^{R*}_{||}\}-e^{-i\phi}\{A^{L}_{0}\widetilde{A}^{L*}_{||}-A^{R}_{0}\widetilde{A}^{R*}_{||}\}]\\
\nonumber&&\qquad
+\frac{m_\ell}{\sqrt{q^2}}
  {\rm Re}[e^{i\phi}\{\widetilde{A}^{L}_{\perp}A^*_S+\widetilde{A}^{R}_{\perp}A^*_S\}-e^{-i\phi}\{A^{L}_{\perp}\widetilde{A}^*_S+A^{R}_{\perp}\widetilde{A}^*_S\}]\Bigg]\\  
s_8&=& -\frac{1}{\sqrt{2}}\beta_\ell^2 {\rm Re}[e^{i\phi}\{\widetilde{A}^{L}_{0}A^{L*}_{\perp}+\widetilde{A}^{R}_{0}A^{R*}_{\perp}\}-e^{-i\phi}\{A^{L}_{0}\widetilde{A}^{L*}_{\perp}+A^{R}_{0}\widetilde{A}^{R*}_{\perp}\}]\\
s_9&=& \beta_\ell^2{\rm Re}[e^{i\phi}\{\widetilde{A}^{L}_{||}A^{L*}_{\perp}+\widetilde{A}^{R}_{||}A^{R*}_{\perp}\}-e^{-i\phi}\{A^{L}_{||}\widetilde{A}^{L*}_{\perp}+A^{R}_{||}\widetilde{A}^{R*}_{\perp}\}]
\end{eqnarray}  
}

The coefficients $h_i$ are given by

\allowdisplaybreaks{
\begin{eqnarray}
h_{1s}&=&\frac{2+\beta_\ell^2}{2}{\rm Re}[e^{i\phi}\{\widetilde{A}^{L}_{\perp}A^{L*}_{\perp}+\widetilde{A}^{L}_{||}A^{L*}_{||}+\widetilde{A}^{R}_{\perp}A^{R*}_{\perp}+\widetilde{A}^{R}_{||}A^{R*}_{||}\}]\\
\nonumber &&\qquad  +\frac{4m_\ell^2}{q^2}{\rm Re}[e^{i\phi}\{\widetilde{A}^{L}_{\perp}A^{R*}_{\perp}+\widetilde{A}^{L}_{||}A^{R*}_{||}\}+e^{-i\phi}\{A^{L}_{\perp}\widetilde{A}^{R*}_{\perp}+A^{L}_{||}\widetilde{A}^{R*}_{||}\}]\\
h_{1c}&=&2{\rm Re}[e^{i\phi}\{\widetilde{A}^{L}_{0}A^{L*}_{0}+\widetilde{A}^{R}_{0}A^{R*}_{0}\}]\\\nonumber
&&\qquad +\frac{8m_\ell^2}{q^2}[{\rm Re}[e^{i\phi}\widetilde{A}_tA_t^*]+{\rm Re}\{e^{i\phi}\widetilde{A}^{L}_0A^{R*}_0+e^{-i\phi}A^{L}_0\widetilde{A}^{R*}_0\}]+2\beta_\ell^2{\rm Re}[e^{i\phi}\{\widetilde{A}_SA_S^*\}]\\
h_{2s}&=&\frac{\beta_\ell^2}{2}{\rm Re}[e^{i\phi}\{\widetilde{A}^{L}_{\perp}A^{L*}_{\perp}+\widetilde{A}^{L}_{||}A^{L*}_{||}+\widetilde{A}^{R}_{\perp}A^{R*}_{\perp}+\widetilde{A}^{R}_{||}A^{R*}_{||}\}]\\
h_{2c}&=&-2\beta_\ell^2{\rm Re}[e^{i\phi}\{\widetilde{A}^{L}_{0}A^{L*}_{0}+\widetilde{A}^{R}_{0}A^{R*}_{0}\}]\\
h_3 &=&\beta_\ell^2{\rm Re}[e^{i\phi}\{\widetilde{A}^{L}_{\perp}A^{L*}_{\perp}-\widetilde{A}^{L}_{||}A^{L*}_{||}+\widetilde{A}^{R}_{\perp}A^{R*}_{\perp}-\widetilde{A}^{R}_{||}A^{R*}_{||}\}]\\
h_4&=&\frac{1}{\sqrt{2}}\beta_\ell^2{\rm Re}[e^{i\phi}\{\widetilde{A}_0^{L}A_{||}^{L*}+\widetilde{A}_0^{R}A_{||}^{R*}\}+e^{-i\phi}\{A_0^{L}\widetilde{A}_{||}^{L*}+A_0^{R}\widetilde{A}_{||}^{R*}\}]\\
h_5&=&\sqrt{2}\beta_\ell\Bigg[{\rm Re}[e^{i\phi}\{\widetilde{A}_0^{L}A_{\perp}^{L*}-\widetilde{A}_0^{R}A_{\perp}^{R*}\}+e^{-i\phi}\{A_0^{L}\widetilde{A}_{\perp}^{L*}-A_0^{R}\widetilde{A}_{\perp}^{R*}\}]\\ \nonumber
&&\qquad -\frac{m_\ell}{\sqrt{q^2}}
   {\rm Re}[e^{i\phi}\{\widetilde{A}^{L}_{||} A_S^*+\widetilde{A}^{R}_{||} A_S^*\}+e^{-i\phi}\{A^{L}_{||} \widetilde{A}_S^*+A^{R}_{||} \widetilde{A}_S^*\}]\Bigg]\\
h_{6s}&=&2\beta_\ell{\rm Re}[e^{i\phi}\{\widetilde{A}_{||}^{L}A_{\perp}^{L*}-\widetilde{A}_{||}^{R}A_{\perp}^{R*}\}+e^{-i\phi}\{A_{||}^{L}\widetilde{A}_{\perp}^{L*}-A_{||}^{R}\widetilde{A}_{\perp}^{R*}\}]\\
h_{6c}&=&4\beta_\ell\frac{m_\ell}{\sqrt{q^2}}{\rm Re}[e^{i\phi}\{\widetilde{A}^L_0 A_S^*+\widetilde{A}^R_0 A_S^*\}+e^{-i\phi}\{A^L_0 \widetilde{A}_S^*+A^R_0 \widetilde{A}_S^*\}]\\
h_7&=&\sqrt{2}\beta_\ell \Bigg[{\rm Im}[e^{i\phi}\{\widetilde{A}_0^{L}A_{||}^{L*}-\widetilde{A}_0^{R}A_{||}^{R*}\}+e^{-i\phi}\{A_0^{L}\widetilde{A}_{||}^{L*}-A_0^{R}\widetilde{A}_{||}^{R*}\}]\\
\nonumber
&&\qquad +\frac{m_\ell}{\sqrt{q^2}}
   {\rm Im}[e^{i\phi}\{\widetilde{A}^{L}_{\perp} A_S^*+\widetilde{A}^{R}_{\perp} A_S^*\}+e^{-i\phi}\{A^{L}_{\perp} \widetilde{A}_S^*+A^{R}_{\perp} \widetilde{A}_S^*\}]\Bigg]\\  
h_8&=&\frac{1}{\sqrt{2}}\beta_\ell^2 {\rm Im}[e^{i\phi}\{\widetilde{A}_0^L A_\perp^{L*}+\widetilde{A}_0^R A_\perp^{R*}\}+e^{-i\phi}\{A_0^L \widetilde{A}_\perp^{L*}+A_0^R \widetilde{A}_\perp^{R*}\}]\\
h_9&=&-\beta_\ell^2 {\rm Im}[e^{i\phi}\{\widetilde{A}_{||}^L A_\perp^{L*}+\widetilde{A}_{||}^R A_\perp^{R*}\}+e^{-i\phi}\{A_{||}^L \widetilde{A}_\perp^{L*}+A_{||}^R \widetilde{A}_\perp^{R*}\}]
\end{eqnarray}
}

In the above expressions, the amplitude $\widetilde{A}_X$ denotes the amplitude $A_X(\bar{B}\to f)$, without applying CP-conjugation to the final state. One has the relation
\begin{equation}
\widetilde{A}_X=\eta_X\bar A_X
\end{equation}
where $\bar A_X$ can be obtained from $A_X$ by changing the sign of all weak phases.



\begin{thebibliography}{99}




\bibitem{1301.1700} 
  {\bf Babar} Collaboration, J.~L.~Ritchie,
  ``Angular Analysis of $B \to K^*\ell^+\ell^-$ in BABAR,''
  arXiv:1301.1700 [hep-ex].


\bibitem{1304.6325} 
  {\bf LHCb} Collaboration,
  ``Differential branching fraction and angular analysis of the decay $B^{0} \to K^{*0} \mu^{+}\mu^{-}$,''
  JHEP {\bf 1308}, 131 (2013),
  arXiv:1304.6325[hep-ex].

\bibitem{1308.1707} 
  {\bf LHCb} Collaboration,
  ``Measurement of Form-Factor-Independent Observables in the Decay $B^{0} \to K^{*0} \mu^+ \mu^-$,''
  Phys.\ Rev.\ Lett.\  {\bf 111}, no. 19, 191801 (2013),
  arXiv:1308.1707 [hep-ex].

\bibitem{1308.3409} 
  {\bf CMS} Collaboration,
  ``Angular analysis and branching fraction measurement of the decay $B^0 \to K^{*0} \mu^+\mu^-$,''
  Phys.\ Lett.\ B {\bf 727}, 77 (2013),
  arXiv:1308.3409 [hep-ex].



\bibitem{1403.8045} 
  {\bf LHCb} Collaboration,
  ``Angular analysis of charged and neutral $B \to K \mu^+\mu^-$  decays,''
  JHEP {\bf 1405}, 082 (2014),
  arXiv:1403.8045 [hep-ex].

\bibitem{1406.6482} 
  {\bf LHCb} Collaboration,
  ``Test of lepton universality using $B^{+}\rightarrow K^{+}\ell^{+}\ell^{-}$ decays,''
  Phys.\ Rev.\ Lett.\  {\bf 113}, no. 15, 151601 (2014),
  arXiv:1406.6482 [hep-ex].



\bibitem{0904.0770} 
  {\bf Belle} Collaboration,
  ``Measurement of the Differential Branching Fraction and Forward-Backword Asymmetry for $B \to K^{(*)}\ell^+\ell^-$,''
  Phys.\ Rev.\ Lett.\  {\bf 103}, 171801 (2009),
  arXiv:0904.0770 [hep-ex].

\bibitem{1108.0695} 
  {\bf CDF} Collaboration,
  ``Measurements of the Angular Distributions in the Decays $B \to K^{(*)} \mu^+ \mu^-$ at CDF,''
  Phys.\ Rev.\ Lett.\  {\bf 108}, 081807 (2012),
  arXiv:1108.0695 [hep-ex].

\bibitem{1204.3933} 
  {\bf Babar} Collaboration,
  ``Measurement of Branching Fractions and Rate Asymmetries in the Rare Decays $B \to K^{(*)} l^+ l^-$,''
  Phys.\ Rev.\ D {\bf 86}, 032012 (2012),
  arXiv:1204.3933 [hep-ex].

\bibitem{1403.8044} 
  {\bf LHCb} Collaboration,
  ``Differential branching fractions and isospin asymmetries of $B \to K^{(*)} \mu^+ \mu^-$ decays,''
  JHEP {\bf 1406}, 133 (2014),
  arXiv:1403.8044 [hep-ex].

\bibitem{1408.0978} 
  {\bf LHCb} Collaboration,
  ``Measurement of $C\!P$ asymmetries in the decays $B^0 \rightarrow K^{*0} \mu^+ \mu^-$ and $B^+ \rightarrow K^{+} \mu^+ \mu^-$,''
  JHEP {\bf 1409}, 177 (2014),
  arXiv:1408.0978 [hep-ex].



\bibitem{1305.2168} 
  {\bf LHCb} Collaboration,
  ``Differential branching fraction and angular analysis of the decay $B_s^0\to\phi\mu^{+}\mu^{-}$,''
  JHEP {\bf 1307}, 084 (2013),
  arXiv:1305.2168 [hep-ex].



\bibitem{1307.5024} 
 {\bf LHCb} Collaboration,
  ``Measurement of the $B^0_s \to \mu^+ \mu^-$ branching fraction and search for $B^0 \to \mu^+ \mu^-$ decays at the LHCb experiment,''
  Phys.\ Rev.\ Lett.\  {\bf 111}, 101805 (2013),
  arXiv:1307.5024 [hep-ex].

\bibitem{1307.5025} 
  {\bf CMS} Collaboration,
  ``Measurement of the $B_s \to \mu^+ \mu^-$ branching fraction and search for $B^0 \to \mu^+ \mu^-$ with the CMS Experiment,''
  Phys.\ Rev.\ Lett.\  {\bf 111}, 101804 (2013),
  arXiv:1307.5025 [hep-ex].

\bibitem{1411.4964} 
  F.~Archilli,
  ``$B^0_{s} \rightarrow \mu^+\mu^-$ at LHC,''
  arXiv:1411.4964 [hep-ex].


\bibitem{1312.5364} 
  {\bf Babar} Collaboration,
  ``Measurement of the $B\to X_s \ell^+ \ell^-$ branching fraction and search for direct CP violation from a sum of exclusive final states,''
  Phys.\ Rev.\ Lett.\  {\bf 112}, 211802 (2014),
  arXiv:1312.5364 [hep-ex].

\bibitem{1402.7134} 
  {\bf Belle} Collaboration,
  ``Measurement of the Lepton Forward-Backward Asymmetry in Inclusive $B \rightarrow X_s \ell^+ \ell^-$ Decays,''
  arXiv:1402.7134 [hep-ex].




\bibitem{1307.5683} 
  S.~Descotes-Genon, J.~Matias and J.~Virto,
  ``Understanding the $B \to K^*\mu^+\mu^-$ Anomaly,''
  Phys.\ Rev.\ D {\bf 88}, no. 7, 074002 (2013),
  arXiv:1307.5683 [hep-ph].

\bibitem{1308.1501} 
  W.~Altmannshofer and D.~M.~Straub,
  ``New physics in $B \to K^*\mu\mu$?,''
  Eur.\ Phys.\ J.\ C {\bf 73}, no. 12, 2646 (2013),
  arXiv:1308.1501 [hep-ph].
  
\bibitem{1310.2478} 
  F.~Beaujean, C.~Bobeth and D.~van Dyk,
  ``Comprehensive Bayesian analysis of rare (semi)leptonic and radiative $B$ decays,''
  Eur.\ Phys.\ J.\ C {\bf 74}, no. 6, 2897 (2014)
  [Erratum-ibid.\ C {\bf 74}, no. 12, 3179 (2014)]
  [arXiv:1310.2478 [hep-ph].
  
\bibitem{1310.3887} 
  R.~R.~Horgan, Z.~Liu, S.~Meinel and M.~Wingate,
  ``Calculation of $B^0 \to K^{*0} \mu^+ \mu^-$ and $B_s^0 \to \phi \mu^+ \mu^-$ observables using form factors from lattice QCD,''
  Phys.\ Rev.\ Lett.\  {\bf 112}, 212003 (2014),
  arXiv:1310.3887 [hep-ph].
  
\bibitem{1408.1627} 
  G.~Hiller and M.~Schmaltz,
  ``$R_K$ and future $b \to s \ell \ell$ physics beyond the standard model opportunities,''
  Phys.\ Rev.\ D {\bf 90}, no. 5, 054014 (2014),
  arXiv:1408.1627 [hep-ph].
  
\bibitem{1408.4097} 
  D.~Ghosh, M.~Nardecchia and S.~A.~Renner,
  ``Hint of Lepton Flavour Non-Universality in $B$ Meson Decays,''
  JHEP {\bf 1412}, 131 (2014),
  arXiv:1408.4097 [hep-ph].
  
 \bibitem{1410.4545} 
  T.~Hurth, F.~Mahmoudi and S.~Neshatpour,
  ``Global fits to b $\to s\ell\ell$ data and signs for lepton non-universality,''
  JHEP {\bf 1412}, 053 (2014),
  arXiv:1410.4545 [hep-ph].
    
\bibitem{1411.3161} 
  W.~Altmannshofer and D.~M.~Straub,
  ``State of new physics in $b\to s$ transitions,''
  arXiv:1411.3161 [hep-ph].

\bibitem{1411.0565} 
  S.~L.~Glashow, D.~Guadagnoli and K.~Lane,
  ``Lepton Flavor Violation in B Decays?,''
  arXiv:1411.0565 [hep-ph].

\bibitem{1412.1791} 
  B.~Gripaios, M.~Nardecchia and S.~A.~Renner,
  ``Composite leptoquarks and anomalies in $B$-meson decays,''
  arXiv:1412.1791 [hep-ph].

\bibitem{1412.7164} 
  B.~Bhattacharya, A.~Datta, D.~London and S.~Shivashankara,
  ``Simultaneous Explanation of the $R_K$ and $R(D^{(*)})$ Puzzles,''
  arXiv:1412.7164 [hep-ph].

  

\bibitem{0904.1869}
  U.~Nierste,
  ``Three Lectures on Meson Mixing and CKM phenomenology,''
  arXiv:0904.1869 [hep-ph].
 
\bibitem{0805.2525}
  C.~Bobeth, G.~Hiller and G.~Piranishvili,
  ``CP Asymmetries in bar $B \to \bar{K}^* (\to \bar{K} \pi) \bar{\ell} \ell$ and Untagged $\bar{B}_s$, $B_s \to \phi (\to K^{+} K^-) \bar{\ell} \ell$ Decays at NLO,''
  JHEP {\bf 0807} (2008) 106,
  arXiv:0805.2525 [hep-ph].
  
\bibitem{1204.1737} 
  K.~De Bruyn, R.~Fleischer, R.~Knegjens, P.~Koppenburg, M.~Merk, A.~Pellegrino and N.~Tuning,
  ``Probing New Physics via the $B^0_s\to \mu^+\mu^-$ Effective Lifetime,''
  Phys.\ Rev.\ Lett.\  {\bf 109}, 041801 (2012),
  arXiv:1204.1737 [hep-ph].
  
\bibitem{1111.4882} 
  S.~Descotes-Genon, J.~Matias and J.~Virto,
  ``An analysis of $B_{d,s}$ mixing angles in presence of New Physics and an update of $B_s \to K^{0*} \bar K^{0*}$,''
  Phys.\ Rev.\ D {\bf 85}, 034010 (2012),
  arXiv:1111.4882 [hep-ph].

  
  
\bibitem{1011.0352} 
  {\bf Belle-II} Collaboration, T.~Abe {\it et al.}
  ``Belle II Technical Design Report,''
  arXiv:1011.0352 [physics.ins-det].

  
  
  
  
  
  

\bibitem{Harrison:1998yr}
  {\bf Babar} Collaboration, P.~F.~Harrison {\it et al.}
  ``The BABAR physics book: Physics at an asymmetric $B$ factory,''
  SLAC-R-0504.

\bibitem{Dunietz:1990cj}
  I.~Dunietz, H.~R.~Quinn, A.~Snyder, W.~Toki and H.~J.~Lipkin,
  ``How to extract CP violating asymmetries from angular correlations,''
  Phys.\ Rev.\ D {\bf 43} (1991) 2193.


\bibitem{Aubert:2004cp}
  {\bf Babar} Collaboration,
  ``Ambiguity-free measurement of $\cos(2\beta)$: Time-integrated and time-dependent angular analyses of $B \to J/\psi K \pi$,''
  Phys.\ Rev.\ D {\bf 71} (2005) 032005
  [hep-ex/0411016].
  

\bibitem{0609037}
  P.~Ball and R.~Zwicky,
  ``Time-dependent CP Asymmetry in $B \to K^* \gamma$ as a (Quasi) Null Test of the Standard Model,''
  Phys.\ Lett.\ B {\bf 642} (2006) 478
  [hep-ph/0609037].

\bibitem{0802.0876}
  F.~Muheim, Y.~Xie and R.~Zwicky,
  ``Exploiting the width difference in $B_s \to \phi \gamma$,''
  Phys.\ Lett.\ B {\bf 664} (2008) 174,
  arXiv:0802.0876 [hep-ph].

\bibitem{0811.1214}
  W.~Altmannshofer, P.~Ball, A.~Bharucha, A.~J.~Buras, D.~M.~Straub and M.~Wick,
  ``Symmetries and Asymmetries of $B \to K^{*} \mu^{+} \mu^{-}$ Decays in the Standard Model and Beyond,''
  JHEP {\bf 0901} (2009) 019,
  arXiv:0811.1214 [hep-ph].


\bibitem{1202.4266}
  J.~Matias, F.~Mescia, M.~Ramon, J.~Virto,
  ``Complete Anatomy of $\bar{B}_d \to \bar{K}^{* 0} (\to K \pi)\ell^+\ell^-$ and its angular distribution,''
  JHEP {\bf 1204} (2012) 104,
  arXiv:1202.4266 [hep-ph].
  

\bibitem{1104.3342}
  S.~Descotes-Genon, D.~Ghosh, J.~Matias and M.~Ramon,
  ``Exploring New Physics in the $\C7$-$\C7'$ plane,''
  JHEP {\bf 1106} (2011) 099,
  arXiv:1104.3342 [hep-ph].
  
\bibitem{1212.2263} 
  S.~J\"ager, J.~Martin Camalich,
  ``On $B \to V \ell \ell$ at small dilepton invariant mass, power corrections, and new physics,''
  JHEP {\bf 1305}, 043 (2013),
  arXiv:1212.2263 [hep-ph].
  
\bibitem{9612313} 
  K.~G.~Chetyrkin, M.~Misiak and M.~Munz,
  ``Weak radiative B meson decay beyond leading logarithms,''
  Phys.\ Lett.\ B {\bf 400}, 206 (1997)
  [Erratum-ibid.\ B {\bf 425}, 414 (1998)]
  [hep-ph/9612313].
  
\bibitem{1411.7677} 
  T.~Huber, M.~Poradzi\'nski and J.~Virto,
  ``Four-body contributions to $ \overline{B}\to {X}_s\gamma $ at NLO,''
  JHEP {\bf 1501}, 115 (2015),
  arXiv:1411.7677 [hep-ph].
  
\bibitem{9907386} 
  F.~Kruger, L.~M.~Sehgal, N.~Sinha and R.~Sinha,
  ``Angular distribution and CP asymmetries in the decays $\bar B \to K^- \pi^+ e^- e^+$ and $\bar B \to \pi^- \pi^+ e^- e^+$,''
  Phys.\ Rev.\ D {\bf 61}, 114028 (2000)
  [Erratum-ibid.\ D {\bf 63}, 019901 (2001)]
  [hep-ph/9907386].
  
\bibitem{1412.7515} 
  {\bf HFAG} Collaboration, Y.~Amhis {\it et al.},
  ``Averages of $b$-hadron, $c$-hadron, and $\tau$-lepton properties as of summer 2014,''
  arXiv:1412.7515 [hep-ex].
  
\bibitem{1005.0571} 
  U.~Egede, T.~Hurth, J.~Matias, M.~Ramon and W.~Reece,
  ``New physics reach of the decay mode $\bar{B} \to \bar{K}^{*0}\ell^+\ell^-$,''
  JHEP {\bf 1010}, 056 (2010),
  arXiv:1005.0571 [hep-ph].
  
\bibitem{1502.00920} 
  L.~Hofer and J.~Matias,
  ``Exploiting the Symmetries of P and S wave for $B \to K^* \mu^+ \mu^-$,''
  arXiv:1502.00920 [hep-ph].
  
\bibitem{0502060} 
  F.~Kruger and J.~Matias,
  ``Probing new physics via the transverse amplitudes of $\bar B^0 \to K^{*0} (\to K^- \pi^+) \ell^+\ell^-$ at large recoil,''
  Phys.\ Rev.\ D {\bf 71}, 094009 (2005)
  [hep-ph/0502060].
  
\bibitem{1006.5013} 
  C.~Bobeth, G.~Hiller and D.~van Dyk,
  ``The Benefits of $\bar{B} \to \bar{K}^* \ell^+ \ell^-$ Decays at Low Recoil,''
  JHEP {\bf 1007}, 098 (2010),
  arXiv:1006.5013 [hep-ph].

\bibitem{1105.0376} 
  C.~Bobeth, G.~Hiller and D.~van Dyk,
  ``More Benefits of Semileptonic Rare B Decays at Low Recoil: CP Violation,''
  JHEP {\bf 1107}, 067 (2011),
  arXiv:1105.0376 [hep-ph].

\bibitem{1106.3283} 
  D.~Becirevic and E.~Schneider,
  ``On transverse asymmetries in $B \to K^* \ell^+\ell^-$,''
  Nucl.\ Phys.\ B {\bf 854}, 321 (2012),
  arXiv:1106.3283 [hep-ph].
  
\bibitem{1207.2753} 
  S.~Descotes-Genon, J.~Matias, M.~Ramon and J.~Virto,
  ``Implications from clean observables for the binned analysis of $B \to K^*\mu^+\mu^-$ at large recoil,''
  JHEP {\bf 1301}, 048 (2013),
  arXiv:1207.2753 [hep-ph].
  
\bibitem{1303.5794} 
  S.~Descotes-Genon, T.~Hurth, J.~Matias and J.~Virto,
  ``Optimizing the basis of ${B} \to {K}^{*}\ell^+ \ell^-$ observables in the full kinematic range,''
  JHEP {\bf 1305}, 137 (2013),
  arXiv:1303.5794 [hep-ph].
  
\bibitem{0106067} 
  M.~Beneke, T.~Feldmann and D.~Seidel,
  ``Systematic approach to exclusive $B \to V \ell^+ \ell^-, V \gamma$ decays,''
  Nucl.\ Phys.\ B {\bf 612}, 25 (2001)
  [hep-ph/0106067].

\bibitem{0412400} 
  M.~Beneke, T.~Feldmann and D.~Seidel,
  ``Exclusive radiative and electroweak $b \to d$ and $b \to s$ penguin decays at NLO,''
  Eur.\ Phys.\ J.\ C {\bf 41}, 173 (2005)
  [hep-ph/0412400].
  
\bibitem{9711280} 
  K.~G.~Chetyrkin, M.~Misiak and M.~Munz,
  ``$\Delta F = 1$ nonleptonic effective Hamiltonian in a simpler scheme,''
  Nucl.\ Phys.\ B {\bf 520}, 279 (1998)
  [hep-ph/9711280].

\bibitem{1407.8526} 
  S.~Descotes-Genon, L.~Hofer, J.~Matias and J.~Virto,
  ``On the impact of power corrections in the prediction of $B \to K^*\mu^+\mu^-$ observables,''
  JHEP {\bf 1412}, 125 (2014),
  arXiv:1407.8526 [hep-ph].
  
\bibitem{9812358} 
  J.~Charles, A.~Le Yaouanc, L.~Oliver, O.~Pene and J.~C.~Raynal,
  ``Heavy to light form-factors in the heavy mass to large energy limit of QCD,''
  Phys.\ Rev.\ D {\bf 60}, 014001 (1999)
  [hep-ph/9812358].
  
\bibitem{0008255} 
  M.~Beneke and T.~Feldmann,
  ``Symmetry breaking corrections to heavy to light B meson form-factors at large recoil,''
  Nucl.\ Phys.\ B {\bf 592}, 3 (2001)
  [hep-ph/0008255].
  
\bibitem{0412079} 
  P.~Ball and R.~Zwicky,
  ``$B_{d,s} \to \rho, \omega, K^*, \phi$ decay form-factors from light-cone sum rules revisited,''
  Phys.\ Rev.\ D {\bf 71}, 014029 (2005)
  [hep-ph/0412079].



\bibitem{1211.1896} 
  A.~J.~Buras, F.~De Fazio and J.~Girrbach,
  ``The Anatomy of Z' and Z with Flavour Changing Neutral Currents in the Flavour Precision Era,''
  JHEP {\bf 1302}, 116 (2013),
  arXiv:1211.1896 [hep-ph].
  
\bibitem{1309.2466} 
  A.~J.~Buras and J.~Girrbach,
  ``Left-handed $Z'$ and $Z$ FCNC quark couplings facing new $b \to s \mu^+ \mu^-$ data,''
  JHEP {\bf 1312}, 009 (2013),
  arXiv:1309.2466 [hep-ph].
  
\bibitem{1412.1446} 
  J.~Brod, A.~Lenz, G.~Tetlalmatzi-Xolocotzi and M.~Wiebusch,
  ``New physics effects in tree-level decays,''
  arXiv:1412.1446 [hep-ph].
  
\bibitem{0205287} 
  A.~Faessler, T.~Gutsche, M.~A.~Ivanov, J.~G.~Korner and V.~E.~Lyubovitskij,
  ``The Exclusive rare decays $B \to K(K^*)\bar{\ell} \ell$ and $B_c \to D(D^*) \bar{\ell} \ell$ in a relativistic quark model,''
  Eur.\ Phys.\ J.\ direct C {\bf 4}, 18 (2002)
  [hep-ph/0205287].
 
 

\bibitem{0404250} 
  B.~Grinstein and D.~Pirjol,
  ``Exclusive rare $B \to K^*\ell^+\ell^-$ decays at low recoil: Controlling the long-distance effects,''
  Phys.\ Rev.\ D {\bf 70}, 114005 (2004)
  [hep-ph/0404250].

\bibitem{1101.5118} 
  M.~Beylich, G.~Buchalla and T.~Feldmann,
  ``Theory of $B \to K^{(*)}\ell^+\ell^-$ decays at high $q^2$: OPE and quark-hadron duality,''
  Eur.\ Phys.\ J.\ C {\bf 71}, 1635 (2011),
  arXiv:1101.5118 [hep-ph].

  



\bibitem{1501.03309} 
  T.~Blake, T.~Gershon and G.~Hiller,
  ``Rare b hadron decays at the LHC,''
  arXiv:1501.03309 [hep-ex].



\bibitem{1205.1438} 
  D.~Das and R.~Sinha,
  ``New Physics Effects and Hadronic Form Factor Uncertainties in $B\to K^* \ell^+ \ell^-$,''
  Phys.\ Rev.\ D {\bf 86}, 056006 (2012),
  arXiv:1205.1438 [hep-ph].


\bibitem{1205.1838} 
  F.~Beaujean, C.~Bobeth, D.~van Dyk and C.~Wacker,
  ``Bayesian Fit of Exclusive $b \to s \bar\ell\ell$ Decays: The Standard Model Operator Basis,''
  JHEP {\bf 1208}, 030 (2012),
  arXiv:1205.1838 [hep-ph].


\bibitem{1205.1845} 
  F.~Mahmoudi, S.~Neshatpour, J.~Orloff,
  ``Supersymmetric constraints from $B_s \to \mu^+\mu^-$ and $B \to K^* \mu^+\mu^-$ observables,''
  JHEP {\bf 1208}, 092 (2012),
  arXiv:1205.1845 [hep-ph].


\bibitem{1205.3683} 
  A.~Y.~Korchin and V.~A.~Kovalchuk,
  ``Contribution of vector resonances to the ${\bar B}_d^0 \to {\bar K}^{*0} \mu^+ \mu^-$ decay,''
  Eur.\ Phys.\ J.\ C {\bf 72}, 2155 (2012),
  arXiv:1205.3683 [hep-ph].


\bibitem{1206.0273} 
  W.~Altmannshofer and D.~M.~Straub,
  ``Cornering New Physics in $b \to s$ Transitions,''
  JHEP {\bf 1208}, 121 (2012),
  arXiv:1206.0273 [hep-ph].


\bibitem{1206.2970} 
  N.~Kosnik,
  ``Model independent constraints on leptoquarks from $b \to s \ell^+ \ell^-$ processes,''
  Phys.\ Rev.\ D {\bf 86}, 055004 (2012),
  arXiv:1206.2970 [hep-ph].


\bibitem{1207.4004} 
  D.~Becirevic and A.~Tayduganov,
  ``Impact of $B\to K^\ast_0 \ell^+\ell^-$ on the New Physics search in $B\to K^\ast \ell^+\ell^-$ decay,''
  Nucl.\ Phys.\ B {\bf 868}, 368 (2013),
  arXiv:1207.4004 [hep-ph].


\bibitem{1209.0262} 
  S.~Descotes-Genon, J.~Matias and J.~Virto,
  ``New Physics constraints from optimized observables in $B\to K^* \ell^+\ell^-$ at large recoil,''
  AIP Conf.\ Proc.\  {\bf 1492}, 103 (2012),
  arXiv:1209.0262 [hep-ph].


\bibitem{1209.1525} 
  J.~Matias,
  ``On the S-wave pollution of $B\to K^* \ell^+\ell^-$ observables,''
  Phys.\ Rev.\ D {\bf 86}, 094024 (2012),
  arXiv:1209.1525 [hep-ph].


\bibitem{1210.5279} 
  T.~Blake, U.~Egede and A.~Shires,
  ``The effect of S-wave interference on the $B^0 \to K^{\ast 0}\ell^+\ell^-$ angular observables,''
  JHEP {\bf 1303}, 027 (2013),
  arXiv:1210.5279 [hep-ph].


\bibitem{1211.6453} 
  T.~Hurth and F.~Mahmoudi,
  ``Colloquium: New physics search with flavor in the LHC era,''
  Rev.\ Mod.\ Phys.\  {\bf 85}, 795 (2013),
  arXiv:1211.6453 [hep-ph].


\bibitem{1212.2321} 
  C.~Bobeth, G.~Hiller and D.~van Dyk,
  ``General analysis of $\bar{B} \to \bar{K}^{(*)}\ell^+ \ell^-$  decays at low recoil,''
  Phys.\ Rev.\ D {\bf 87}, no. 3, 034016 (2013),
  arXiv:1212.2321 [hep-ph].



\bibitem{1301.7535} 
  H.~Gong, Y.~D.~Yang and X.~B.~Yuan,
  ``Constraints on anomalous $tcZ$ coupling from $\bar B \to \bar K^* \mu^+ \mu^-$ and $B_s \to \mu^+ \mu^-$ decays,''
  JHEP {\bf 1305}, 062 (2013),
  arXiv:1301.7535 [hep-ph].


\bibitem{1305.4808} 
  S.~Descotes-Genon, T.~Hurth, J.~Matias and J.~Virto,
  ``$B \to K^* \ell\ell$: The New Frontier of New Physics searches in Flavor,''
  arXiv:1305.4808 [hep-ph].


\bibitem{1306.3775} 
  A.~J.~Buras and J.~Girrbach,
  ``Towards the Identification of New Physics through Quark Flavour Violating Processes,''
  Rept.\ Prog.\ Phys.\  {\bf 77}, 086201 (2014)
  arXiv:1306.3775 [hep-ph].




\bibitem{1308.1959} 
  R.~Gauld, F.~Goertz, U.~Haisch,
  ``On minimal $Z'$ explanations of the $B\to K^*\mu^+\mu^-$ anomaly,''
  Phys.\ Rev.\ D {\bf 89}, no. 1, 015005 (2014),
  arXiv:1308.1959 [hep-ph].


\bibitem{1308.4379} 
  C.~Hambrock, G.~Hiller, S.~Schacht, R.~Zwicky,
  ``$B \to K^\star$ form factors from flavor data to QCD and back,''
  Phys.\ Rev.\ D {\bf 89}, 074014 (2014),
  arXiv:1308.4379 [hep-ph].




\bibitem{1310.1082} 
  R.~Gauld, F.~Goertz and U.~Haisch,
  ``An explicit $Z'$-boson explanation of the $B \to K^* \mu^+ \mu^-$ anomaly,''
  JHEP {\bf 1401}, 069 (2014),
  arXiv:1310.1082 [hep-ph].


\bibitem{1310.1937} 
  A.~Datta, M.~Duraisamy and D.~Ghosh,
  ``Explaining the $B \to K^\ast \mu^+ \mu^-$ data with scalar interactions,''
  Phys.\ Rev.\ D {\bf 89}, no. 7, 071501 (2014),
  arXiv:1310.1937 [hep-ph].



\bibitem{1310.3722} 
  R.~R.~Horgan, Z.~Liu, S.~Meinel and M.~Wingate,
  ``Lattice QCD calculation of form factors describing the rare decays $B \to K^* \ell^+ \ell^-$ and $B_s \to \phi \ell^+ \ell^-$,''
  Phys.\ Rev.\ D {\bf 89}, no. 9, 094501 (2014),
  arXiv:1310.3722 [hep-lat].



\bibitem{1311.3876} 
  S.~Descotes-Genon, J.~Matias and J.~Virto,
  ``Optimizing the basis of $B\to K^{*}l^{+}l^{-}$ observables and understanding its tensions,''
  PoS EPS {\bf -HEP2013}, 361 (2013),
  arXiv:1311.3876 [hep-ph].



\bibitem{1311.6729} 
  A.~J.~Buras, F.~De Fazio and J.~Girrbach,
  ``331 models facing new $b \to s\mu^+ \mu^-$ data,''
  JHEP {\bf 1402}, 112 (2014),
  arXiv:1311.6729 [hep-ph].



\bibitem{1312.1923} 
  G.~Hiller and R.~Zwicky,
  ``(A)symmetries of weak decays at and near the kinematic endpoint,''
  JHEP {\bf 1403}, 042 (2014),
  arXiv:1312.1923 [hep-ph].


\bibitem{1312.5267} 
  T.~Hurth and F.~Mahmoudi,
  ``On the LHCb anomaly in B $\to K^*\ell^+\ell^-$,''
  JHEP {\bf 1404}, 097 (2014),
  arXiv:1312.5267 [hep-ph].



\bibitem{1401.2145} 
  F.~Mahmoudi, S.~Neshatpour and J.~Virto,
  ``$B \to K^{*} \mu^{+} \mu^{-}$ optimised observables in the MSSM,''
  Eur.\ Phys.\ J.\ C {\bf 74}, no. 6, 2927 (2014),
  arXiv:1401.2145 [hep-ph].




\bibitem{1401.6707} 
  M.~Ahmady, R.~Campbell, S.~Lord and R.~Sandapen,
  ``Predicting the $B \to K^*$ form factors in light-cone QCD,''
  Phys.\ Rev.\ D {\bf 89}, no. 7, 074021 (2014),
  arXiv:1401.6707 [hep-ph].




\bibitem{1402.2844} 
  G.~Isidori and F.~Teubert,
  ``Status of indirect searches for New Physics with heavy flavour decays after the initial LHC run,''
  Eur.\ Phys.\ J.\ Plus {\bf 129}, 40 (2014),
  arXiv:1402.2844 [hep-ph].




\bibitem{1402.6855} 
  J.~Matias and N.~Serra,
  ``Symmetry relations between angular observables in $B^0 \to K^* \mu^+\mu^-$ and the LHCb $P_5^\prime$ anomaly,''
  Phys.\ Rev.\ D {\bf 90}, no. 3, 034002 (2014),
  arXiv:1402.6855 [hep-ph].


\bibitem{1403.1269} 
  W.~Altmannshofer, S.~Gori, M.~Pospelov and I.~Yavin,
  ``Quark flavor transitions in $L_\mu-L_\tau$ models,''
  Phys.\ Rev.\ D {\bf 89}, no. 9, 095033 (2014),
  arXiv:1403.1269 [hep-ph].


\bibitem{1403.2944} 
  P.~Biancofiore, P.~Colangelo and F.~De Fazio,
  ``Rare semileptonic $B\to K^* \ell^+ \ell^- $ decays in RS$_c$ model,''
  Phys.\ Rev.\ D {\bf 89}, no. 9, 095018 (2014),
  arXiv:1403.2944 [hep-ph].




\bibitem{1405.3850} 
  A.~J.~Buras, F.~De Fazio and J.~Girrbach-Noe,
  ``$Z$-$Z'$ mixing and $Z$-mediated FCNCs in SU(3)$_{C}$  x  SU(3)$_{L}$  x U(1)$_{X}$ models,''
  JHEP {\bf 1408}, 039 (2014),
  arXiv:1405.3850 [hep-ph].


\bibitem{1405.5182} 
  W.~Altmannshofer,
  ``New Physics Interpretations of the $B \to K^*\mu^+\mu^-$ Anomaly,''
  arXiv:1405.5182 [hep-ph].



\bibitem{1406.0566} 
  J.~Lyon and R.~Zwicky,
  ``Resonances gone topsy turvy - the charm of QCD or new physics in $b \to s \ell^+ \ell^-$?,''
  arXiv:1406.0566 [hep-ph].




\bibitem{1407.6700} 
  M.~R.~Ahmady, S.~Lord and R.~Sandapen,
  ``Isospin asymmetry in $B \to K^* \mu^+ \mu^-$ using AdS/QCD,''
  Phys.\ Rev.\ D {\bf 90}, no. 7, 074010 (2014),
  arXiv:1407.6700 [hep-ph].





\bibitem{1411.0131} 
  S.~Sun,
  ``Little Flavor: Heavy Leptons, $Z'$ and Higgs Phenomenology,''
  arXiv:1411.0131 [hep-ph].


\bibitem{1411.0922} 
  S.~Descotes-Genon, L.~Hofer, J.~Matias and J.~Virto,
  ``QCD uncertainties in the prediction of $B \to K^* \mu^+ \mu^-$ observables,''
  arXiv:1411.0922 [hep-ph].




\bibitem{411.6423} 
  H.~B.~Fu, X.~G.~Wu and Y.~Ma,
  ``$B\to K^*$ Transition Form Factors and the Semi-leptonic Decay $B \to K^* \mu^+ \mu^-$,''
  arXiv:1411.6423 [hep-ph].



\bibitem{B1412.1003} 
  M.~Blanke,
  ``Flavour Physics Beyond the Standard Model: Recent Developments and Future Perspectives,''
  arXiv:1412.1003 [hep-ph].



\bibitem{1412.2955} 
  G.~Kumar and N.~Mahajan,
  ``$B\rightarrow K^{*}l^+ l^-$: Zeroes of angular observables as test of standard model,''
  arXiv:1412.2955 [hep-ph].


\bibitem{1412.3183} 
  S.~J\"ager and J.~Martin Camalich,
  ``Reassessing the discovery potential of the $B \to K^{*} \ell^+\ell^-$ decays in the large-recoil region: SM challenges and BSM opportunities,''
  arXiv:1412.3183 [hep-ph].



\bibitem{1501.00367} 
  R.~R.~Horgan, Z.~Liu, S.~Meinel and M.~Wingate,
  ``Rare $B$ decays using lattice QCD form factors,''
  arXiv:1501.00367 [hep-lat].



\bibitem{1501.00993} 
  A.~Crivellin, G.~D'Ambrosio and J.~Heeck,
  ``Explaining $h\to\mu^\pm\tau^\mp$, $B\to K^* \mu^+\mu^-$ and $B\to K \mu^+\mu^-/B\to K e^+e^-$ in a two-Higgs-doublet model with gauged $L_\mu-L_\tau$,''
  arXiv:1501.00993 [hep-ph].




\bibitem{1501.05193} 
  S.~Sahoo and R.~Mohanta,
  ``Scalar leptoquarks and the rare B meson decays,''
  arXiv:1501.05193 [hep-ph].
  
\bibitem{1501.03038} 
  {\bf LHCb} Collaboration,
  ``Angular analysis of the $B^0 \rightarrow K^{*0} e^+ e^-$ decay in the low-$q^2$ region,''
  arXiv:1501.03038 [hep-ex].


  
  
\end{thebibliography}
\end{document}